\documentclass[10pt]{iopart}
\usepackage{iopams}
\usepackage{graphicx}
\usepackage{calrsfs}
\usepackage{cite}
\renewcommand{\vec}[1]{\mbox{\protect\boldmath$#1$}}
\newcommand{\gconst}{}
\newcommand{\rd}{{\rm{}d}}
\newcommand{\rh}{r_{\!\mbox{\tiny{}H}}}
\newcommand{\rs}{r_{\!\mbox{\tiny{}S}}}

\newcommand{\rmsv}{\sigma_{\star}}
\newcommand{\rmhor}{\mbox{\tiny{}H}}
\newcommand{\mbh}{M_{\bullet}}
\newcommand{\jbh}{J_{\bullet}}
\newcommand{\jd}{J_{\rm{}d}}
\newcommand{\md}{M_{\rm{}d}}
\newcommand{\mst}{M_\star}
\newcommand{\sun}{\odot}

\newcommand{\wsg}{{\sc wsg}}
\newcommand{\ssg}{{\sc ssg}}
\newcommand{\gr}{{\sc gr}}
\newcommand{\bh}{{\sc bh}}
\newcommand{\isl}{{\sc is}}
\newcommand{\Psid}{\nu_{\rm d}}
\newcommand{\barPsid}{\langle{\nu}_{\rm d}\rangle}
\newcommand{\barPsidn}{\langle{\nu}_{\rm d}\rangle_{0}}
\newcommand{\Psih}{\nu_{\bullet}}
\newcommand{\sezer}[1]{}

\begin{document}
\topical[Topical review]{Gravitating discs around black holes}
\author{V~Karas,$^{\!1,2}$ J-M~Hur\'e$^{3,4}$ and O~Semer\'ak$^2$}
\address{$^1$\,Astronomical Institute, Academy of Sciences of the Czech 
 Republic, Bo\v{c}n\'{\i}~II, CZ-141\,31~Prague, Czech Republic}
\address{$^2$\,Faculty of Mathematics and Physics, Charles~University,
 V~Hole\v{s}ovi\v{c}k\'ach~2, CZ-180\,00~Prague, Czech~Republic}
\address{$^3$\,Observatoire de Paris-Meudon, 
 Place Jules Janssen, F-92195~Meudon, France}
\address{$^4$\,Universit\'e~Paris~7 (Denis Diderot), 2~Place Jussieu,
 F-75251~Paris, France}
\begin{abstract}
Fluid discs and tori around black holes are discussed within different
approaches and with the emphasis on the role of disc gravity. First
reviewed are the prospects of investigating the gravitational field of a
black hole--disc system by analytical solutions of stationary, axially
symmetric Einstein's equations. Then, more detailed considerations are
focused to middle and outer parts of extended disc-like configurations
where relativistic effects are small and the Newtonian description is
adequate.

Within general relativity, only a static case has been analysed in
detail. Results are often very inspiring, however, simplifying 
assumptions must be imposed: ad hoc profiles of the disc density are
commonly assumed and the effects of frame-dragging and completely
lacking. Astrophysical discs (e.g.\ accretion discs in active galactic
nuclei) typically extend far beyond the relativistic domain and are
fairly diluted. However, self-gravity is still essential for their
structure and evolution, as well as for their radiation emission and the
impact on the environment around. For example, a nuclear star cluster in
a galactic centre may bear various imprints of mutual star--disc
interactions, which can be recognised in observational properties, such
as the relation between the central mass and stellar velocity
dispersion.
\end{abstract}
\submitto{\CQG}
\renewcommand{\fnsymbol}{^{\arabic{footnote}}\sezer}
\setcounter{footnote}{0}

\section{Introduction}
The interplay between gravitational attraction and non-gravitational
forces produces systems with various shapes in the universe, including
elongated structures, spheroids, discs, shells and rings. All of them
participate in variety of processes and act on different scales, ranging
from cosmological dimensions down to planetary sizes. Some
configurations, spheroidal in particular, hold by themselves, while
other are maintained by an imposed control of a central body or an
external field. This paper  deals with discs and tori --- typical
configurations of rapidly circulating fluids that are governed by a
central object. Its gravitational attraction supports global stability
of the configuration, at least below a certain critical density of the
disc,  while compactness of the centre suggests prospective 
observational methods. High density configurations suffer from
instabilities and their mass contributes to general-relativity effects 
jointly with the central mass.

Nowadays it should be self-evident that black-hole discs are of
indisputable relevance to different research communities, ranging from
pure relativists to astro-physicists and observational astronomers.
This diversification is reflected in varied approaches that people
adopt to the problem. A mathematically minded theorist would take more
interest in non-trivial  (but still manageable) analytical solutions to
Einstein's equations endowed with axial symmetry. An observer will want
to actually {\it{}see} gas flows swirling in an accretion disc near a 
black hole (\bh) or a compact star. Astronomers often invoke accretion 
discs to explain spectral features observed in galactic nuclei and in
close (interacting) binary stars, although detailed physics of accretion 
flows still remains uncertain. Therefore, it is tempting to search for
common intersections in such a broad subject. However, we will argue
that serious distinctions cannot be neglected which prevent one to
develop an  all-encompassing scheme that would describe inherently
different configurations. In particular, effects of general relativity
(\gr)  have a chance to govern only the innermost regions of accretions
discs and their role is quickly diminished with distance from the
centre. Local (non-gravitational) physics plays crucial role and it 
often dominates over gravitation and governs observable properties of
real systems (perhaps the most profound example arises if one attempts
to compare accretion discs around stars with those around super-massive
{\bh}s). In order to see where limitations of any unified scheme stem
from, it is useful to discuss both simplified and general analytical
models  and some particular astronomical objects in more detail.

In this paper, two-dimensional models are discussed. Needles to say,
this is already enormous restriction. Obtained solutions offer a
convenient tool to elucidate physical mechanisms connected with
non-spherical sources of strong gravitational field. Compact toroids
represent a non-trivial model of such objects that require {\gr}, although
Newtonian approach provides a sufficiently accurate description in many 
cases. Interaction with the central field is especially 
intriguing if the latter is generated by an ultra-compact body. 

The paper starts with a summary of astrophysical motivations
(section~\ref{sec:motivations}). Our attention is focused on discs
around super-massive black holes, because in this case the importance of
self-gravity has been traditionally  acknowledged, but we also mention
compact binary stars, in which case it has been recognised relatively
recently that self-gravity may also play a role, especially in the
context of neutron star collisions, tidal disruption and subsequent
emission of high-energy gamma-rays and, possibly, gravitational waves.
In section~\ref{sec:relativity}, the problem of massive discs around
black holes is discussed within {\gr}. Whereas {\gr} is mainly relevant
for the inner parts of self-gravitating discs (within one hundred
Schwarzschild radii, $\lesssim10^2\rs$), the  field of actual discs is
equally important for their structure at larger distances from the
centre where the Newtonian regime is adequate (section
\ref{sec:astrophysical}). Evidence for self-gravity in accretion discs
is still rather tentative and incomplete; some possibilities and
observational implications are explored in section
\ref{sec:consequences}. There we also indicate the potential role of
other ingredients, such as a compact nuclear star cluster in which the
accretion flow may be embedded. However, the reader should be aware that
important parts of the whole debate remain untouched in the present
paper. In particular, we do not discuss magneto-hydrodynamical
processes, neither do we address non-axially  symmetric modes of the
disc flow. One should bear in mind that these processes are probably
crucial for understanding the mechanism of  viscosity as well as
launching of jets in astronomical objects.

Disc structures are ubiquitous in such objects where the fluid
circulates around and inflows onto a central compact body.  The central
mass, $\mbh$, can be considered as one of the model parameters and
varied by many orders.\footnote{We introduce $\mbh$ for the black-hole mass
in order to maintain clear distinction from the disc mass. Both 
contributions add to the total mass of the system.
Similar situation arises with angular momentum, as discussed later in the 
text.} Its value provides primary classification of
cosmic objects. Physical characteristics of accretion discs scale
roughly with $\mbh$. Indeed, accretion discs around super-massive black
holes in active galactic nuclei and quasars share {\it{}some} properties
with circumstellar discs in close binary systems (e.g., cataclysmic
variable stars, X-ray binaries and micro-quasars).  For example, both
the discs around stellar-mass black holes and accretion discs around
super-massive black  holes may develop local and global instabilities
which involve gravity. However, there are important distinctions between
these two kinds of objects which prohibit any simplistic scaling. It
turns out that galactic nuclear discs tend to be cooler and less dense
compared to circumstellar discs, which translates to different opacity
of the material, its ionization structure and, hence, different spectral
properties. There has been also a very lively and still open discussion,
whether the presence of a black hole in stellar-mass {\bh} candidates
(or, rather, the absence of any hard surface) can be revealed in the
radiation properties of their discs.  

Environmental effects bring more distinctions between different objects
with accretion discs. Unfortunately, at this point a bias is always
inevitably introduced to our models by assumptions about turbulent
transport of matter and momentum. Most solutions are model dependent.
Neglecting this dependence can  hardly be avoided when showing results
of computations,  nevertheless, we attempt to accentuate generic
features when possible.

We refer to monographs by Frank, King \& Raine \cite{FraKR02} and Kato,
Fukue \& Mineshige \cite{KatFM98} for exposition of accretion theory, to
Krolik \cite{Kro99} and Peterson \cite{Pet97} for discussion of
accretion flows with the emphasis on active galactic nuclei, and to
various review papers \cite{AbrP97,Colea02,Kin03,LinP96,PapL95}
summarizing recent advances as well as the problems encountered by
traditional scenarios. 

\section{Astrophysical motivations and some early works}
\label{sec:motivations}
Accretion plays a crucial role in the processes of energy liberation and
mass accumulation that take place in the core of astronomical objects.
Disc-type accretion represents an important mode which is realized under
suitable circumstances, defined by the global geometrical arrangements,
local microphysics of the fluid medium and the initial  conditions as
well. Rotation is crucial and it puts discs, toroids and rings in the
same category of centrifugally supported systems. By accretion discs one
understands topologically toroidal fluid configurations orbiting around
a central body and gradually drifting towards it, provided a mechanism
to  redistribute the angular momentum. The centre governs or at least
substantially contributes to the gravitational field in which the medium
evolves. The formation of the central body is presumed as being
completed, although the accretion process gradually transfers mass and
angular momentum onto the central body and thus contributes to its
subsequent evolution. 

It is worth noting that the genesis of our own solar system  with a star
in the centre was imagined, by Pierre-Simon Laplace and Immanuel Kant,
as emerging out of a turbulent, slowly spinning nebula with flattened,
disc-like shape.\footnote{``{\ldots}~We see a region of space extending
from the centre of the Sun to unknown distances, contained between two
planes not far from each other~{\ldots}'' (I~Kant, {\it{}Allgemeine
Naturgeschichte und Theorie des Himmels}, 1755). Independently,  P~S
Laplace advanced and published a theory of Saturn's rings and the
nebular hypothesis of the planetary system ({\it M\'emoire sur la
th\'eorie de l'anneau de Saturne}, M\'em.\ Acad.\ Sci., 1789;
{\it{}Exposition du systeme du monde}, 1796).} This was in the final
years of the 18th century. The potential associated with a gravitating
ring was examined in detail by F~W Dyson \cite{Dys93}. Later,
considerations associated with star and planet formation were the main 
driving force behind the initial development of accretion disc theory.
Modern evidence for such configurations was reported as early as in
1930s by Baxandall \cite{BaxNS30} in the context of binary stars.  In
the attempt to interpret spectrograms of the famous $\beta$~Lyrae binary
system and in order to explain the  origin of secondary lines, visibly
displaced in their wavelength, ``circulating currents in the outer
envelope'' were tentatively invoked together with a gaseous envelope
that surrounds the larger component. Since then,  interacting binary
stars have been the prototypical systems for studying accretion discs.

The treatment of self-gravitating discs was pursued in Ostriker's 
Newtonian equilibria of uniformly rotating, polytropic, slender rings
\cite{Ost64}. Later, self-gravitating configurations with  realistic
equation of state and opacity were constructed and the basic  formalism
for self-gravitating {\bh} discs was given in  {\gr}
\cite{Bar73b,Boy65,FisM76}. In stellar binary systems, the gas in the
disc is provided by a donor star via Roche-lobe overflow or stellar wind
\cite{Kop78}.  Accretion process helps to feed the central body. If
conditions are suitable, the gas radiates and provides a way to observe
and study the system. Historically, limiting cases were distinguished
for methodological reasons and for the sake of simplification. When
speaking of hydrodynamical non-self-gravitating accretion, the main
assumptions concern stationarity, spatial symmetries that are imposed on
the flow and on the gravitational field (spherical flows with negligible
angular momentum, axially symmetric accretion onto a moving body, and
disc-like planar accretion), and the sound speed profile (subsonic
versus supersonic flows). See, e.g.\ \cite{Bon52,FonI98,HoyL39,HoyL41}
and references cited therein. It turns out that behaviour of truly
three-dimensional unsteady flows is rather different and more complex
\cite{VilH03,IguNA03}.

Nearly spherical accretion occurs if the angular momentum content of
accreted material is negligible (this case appears appropriate mainly for
accretion onto an isolated body). In another limit, one ignores radial
transport of the material and considers an axially symmetric toroidal
configuration that preserves perfect rotation about the symmetry axis.
Both these cases   were treated in many works, assuming the test-fluid
approximation (flow  lines are determined by the central gravity and
pressure gradients only). Toroidal topology of the fluid system
introduces the main difference when compared with the physics of
rotating stars (for the latter, see a text-book exposition of the
subject by Tassoul \cite{Tas78}). On large scales, the accretion
mechanism was proposed as the energy source in bright galactic nuclei in
late 1960s (Lynden-Bell \cite{Lyn69}). Since then the idea has been
continuously developed  and brought to a text-book level.

Self-gravity has global consequences on the disc shape, namely, 
it influences the location of its inner and outer edges, as well as  
the disc geometrical thickness. The effect
can be roughly parameterised by the ratio of the disc mass to the
central {\bh} mass, $q\equiv\md/\mbh$ \cite{AbrCCW84,KozJA78}. 
Wilson \cite{Wil81} speculated about the importance of 
self-gravity (especially in cataclysmic binary systems), 
which would have direct observational
consequences, but this possibility was later rejected on the basis
of more detailed arguments; it turns out that non-standard (massive)
discs are ruled out in cataclysmic binaries
\cite{HubHS94}. Self-gravity is normally unimportant 
in binary star systems also because their discs have limited space to occupy.
Under typical  conditions (i.e.\ among well-known cosmic systems), 
the disc mass plays a role in the process of gradual 
fragmentation of proto-planetary nebulae, formation of planetary bodies, 
and their subsequent migration. (Recent observations, in
particular with the Hubble Space Telescope, have demonstrated that discs
are a by-product of star formation and also present during the birth of
planets.) Also planetary rings are subject of self-gravitational
instabilities --- but we cannot tackle all these lively topics here 
(see \cite{GreB84,Lar97,RicABB03,War97}).
There has been a recent interest in high-density (neutron) tori,
in which case the disc own gravity must be important. This kind of massive discs 
could be formed in the course of binary star coalescence or as
a product of collapse to a {\bh} \cite{HachEN86,NakOK87}. 
The question of stability and time evolution of these heavy, transient 
tori is essential.
The idea was advanced in the framework of scenarios of failed 
supernovae and strongly magnetised tori
\cite{Pac98,Broetal00,MacFW99,vPut01,vPutL03}, and an 
interesting possibility of 
intermittent accretion was hypothesised in the context of massive neutron 
tori that may be formed from a tidally disrupted compact star 
\cite{ZanRF03}. 

As mentioned above, gravitational field of heavy discs influences the 
global structure of the system, which in turn determines possible 
figures of equilibrium. Coming back to accretion discs in galactic 
nuclei, vertical structure of these discs is affected by their 
self-gravity especially in the middle and the outer
regions (beyond a few hundreds $\rs$, say) 
where the vertical component of the disc gravitational field exceeds
the corresponding $z$ component of the central field
\cite{ShlB87,ShlB89,ShlBF90,Sto93}. The main aspect of
discs' non-negligible gravitational effect on their own structure
concerns the stability with respect to perturbations. Self-gravity
was first examined in the context of accretion discs around
super-massive black holes by Paczy\'nski \cite{Pac78a}, who
proposed that local (Jeans-type) self-gravitational instabilities are
invoked if the density exceeds a threshold value. Turbulent 
fragmentation may contribute to viscosity and to heating of
the disc medium. Even the formation
of stars may be induced in this way \cite{KozWP79,ShoW82}. 
Collin \& Zahn \cite{ColZ99,ColZ99a} explored the possibility and
the impact of star-formation in galactic-centre discs.
Recently, also Sirko \& Goodman \cite{SirG03} discussed consequences which
the formation of massive stars in the disc may have on its heating and spectrum.
Non-axisymmetric instabilities of accretion discs and tori have been 
widely investigated (see e.g.\ \cite{Bla87,GooN88,PapP84} and
references cited therein). Their importance for modelling
realistic accretion discs has been soon recognised, but we do not 
examine these types of instability
here. If one considers strictly axisymmetric and stationary rotational motion,
one finds idealized configurations, which provide useful insight
into the problem of accretion discs; for example, the question of
the role of 
mean stationary solutions. However, there is no doubt that these
solutions must be rather different from real 
accretion flows; investigations of the latter has culminated in recent 
three-dimensional numerical simulations \cite{VilH03}.
It is noteworthy
here that we are allowed to treat both accretions discs in 
binary stars (with an accreting stellar-mass black hole) and 
in galactic nuclei, because common physical mechanisms operate
in both cases. However, there are important distinctions which 
become particularly essential when realistic configurations are discussed.
We will point out limitations of simple scaling between the two limiting
cases later in the text, but great part
of this paper concerns very simplified situations in which rough scaling
according to the system size and mass is possible.
 
\begin{figure*}[tb]
\begin{center}
\includegraphics[width=0.45\textwidth]{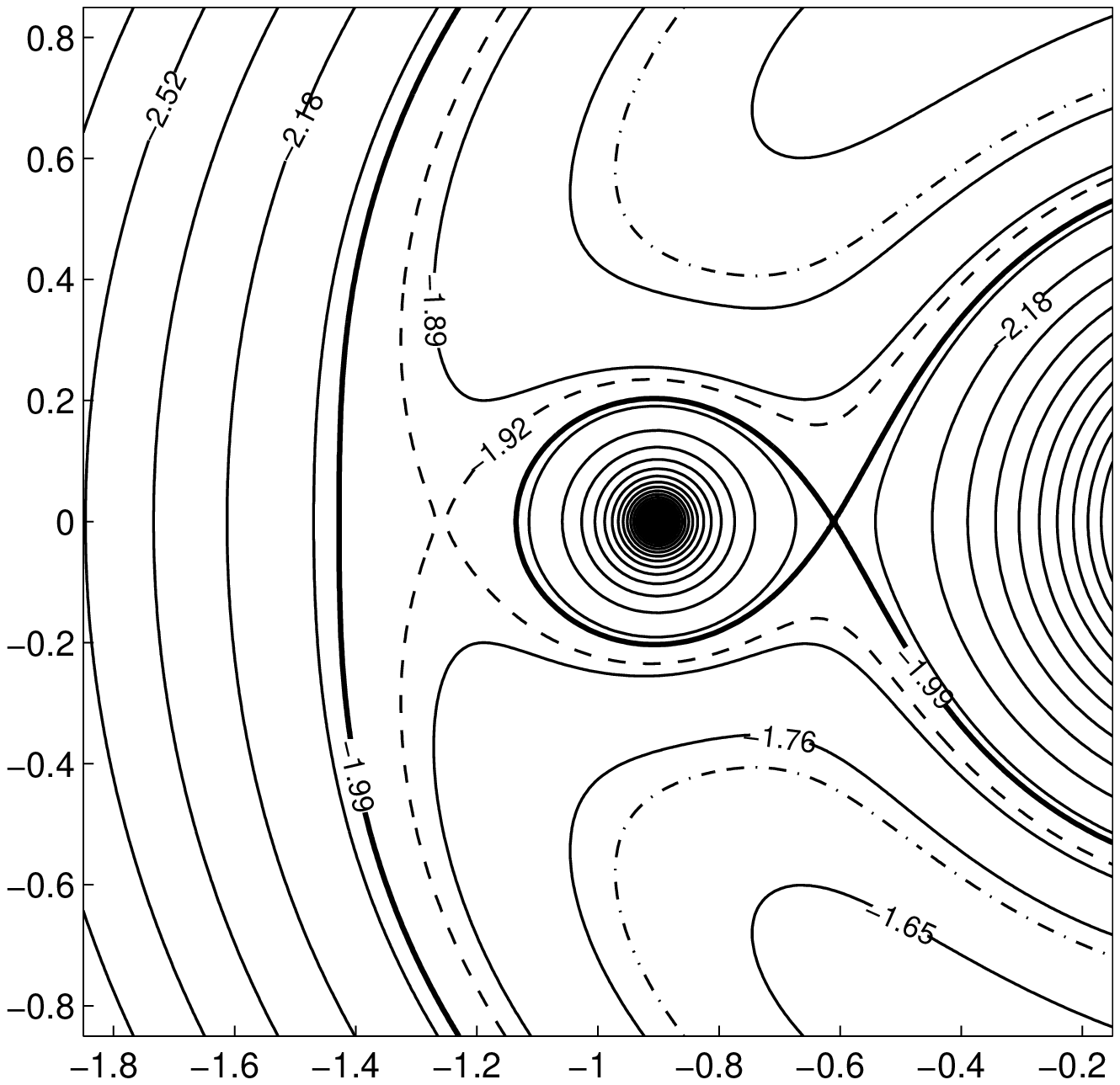}
\hspace*{5mm}
\includegraphics[width=0.45\textwidth]{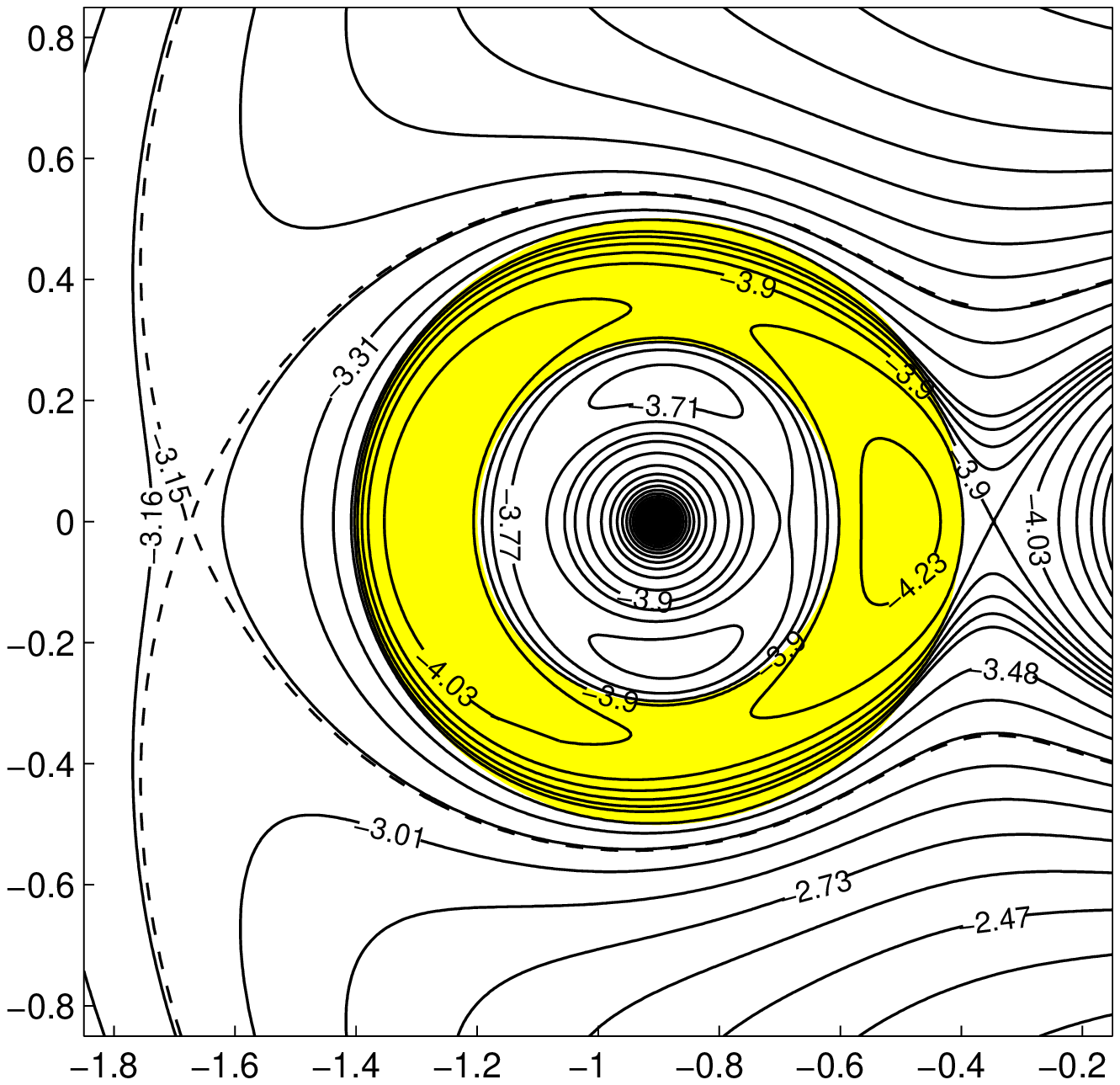}
\end{center}
\caption{Left: Contours of the Roche potential are plotted for two point
masses (``stars'') in their orbital plane. Here, the ratio of masses is 
assumed to be 0.1 (coordinates are scaled by separation between the two
components; no disc is present). Right: The same type of graph as on the left 
side, but with a contribution of a thin planar disc taken into account 
as an additional source of the gravitational field. 
The plot shows the potential structure in the critical region near 
to the cusp, where the mass exchange takes place. See \cite{Wil81}
for an original discussion of the binary Roche potential with 
a massive disc taken into account.}
\label{fig:v0}
\end{figure*}

In the next section we summarize techniques that have been employed to
derive spacetimes of gravitating discs around black holes within \gr\ 
framework and then we proceed to possible astrophysical
applications of massive discs. We start the discussion by
briefly recalling a uncomplicated case of a light non-self-gravitating 
Newtonian disc that
can be formed at a certain evolutionary stage of a binary star when the
donor component dumps material on the other one. A text-book analysis
of matter transfer employs the notion of Roche potential in which
the actual gravitational field of a binary is approximated by
the field of two massive points orbiting around a common centre of 
mass \cite{BoyBKCh02} (see the left panel of figure~\ref{fig:v0}). An
accretion disc forms when the donor fills its Roche lobe and starts
overflow near the Lagrange L$_1$ point. The gas transferred in this way
possesses high angular momentum and creates a disc-like structure before
being captured by the accreting body. The Roche model gives a basic
classification of binary stars into three main categories according to the
sizes of individual components: detached, semi-detached and contact
systems. As mentioned above, their discs can usually be treated as 
non-gravitating. 

In contrast to this transparent scheme, a sufficiently massive disc could
change the potential field near the cusp. In the right panel of
figure~\ref{fig:v0}, distortion of equipotential surfaces is depicted with 
the disc contribution to the gravitational field taken into account
(here, the disc mass is comparable to that of the secondary star).
It is relevant to remark that equilibria
of relativistic tori also show the structure of critical lobes, 
which is similar to the Roche geometry, 
even if the fluid is non-self-gravitating 
and the only source of gravity is the central black hole (see
section~\ref{testfluid}). This structure facilitates accretion onto
the central body.

Equilibrium figures of critical (lobe-filling) tori 
represent a starting point for the discussion of accreting black 
holes in compact binary systems. Naturally, the Roche lobe geometry
is applicable for the mass overflow onto a black hole as much
as it is in other systems of close binaries, in which the mass
is exchanged between the components. However, topologically similar 
configurations of critical tori can arise also when accretion
proceeds onto an isolated black hole. In this case the 
torus can develop a cusp and the overflow for a purely relativistic 
reason. The latter 
situation is particularly important for accretion onto super-massive
{\bh}s in galactic nuclei.

An issue of global stability of such configurations must be raised: 
if a small amount of material is transferred
through the inner cusp onto the central body, then a new, modified
distribution of mass and of angular momentum is established. Geometrical
shape of the critical equilibrium is changed accordingly. A
question arises whether the new configuration is stable or whether,
instead, it undergoes further accretion in a catastrophically
increasing (runaway) rate. Abramowicz {\etal} \cite{AbrCN83}
claimed the presence of this kind of instability in pseudo-Newtonian
massive tori, while Wilson \cite{Wil84} demonstrated that
mass-less tori (i.e.\ test fluid) around a Kerr {\bh} are not
run-away unstable. Answer to the problem of this
instability depends on the radial distribution of angular momentum
throughout the initial configuration (before the mass
exchange takes place) and, in the relativistic case, also on {\bh} 
angular momentum parameter. It was demonstrated that increasing {\bh} 
rotation and a positive slope of the angular momentum distribution 
both have a stabilizing effect
\cite{AbrKL98,DaiM97,FonD02,LuChYZ00}. 
Increasing $\md$, on the other hand, acts against stability 
\cite{NisLEA96,MasE97,MasNE98,NisE96}. 
Comparisons \cite{FonD03} are not easy because different approaches
have been employed in papers (numerical techniques vs.\ analytical calculations;
proper \gr\ treatments vs.\ approximations and
pseudo-Newtonian models).
The black-hole--disc systems are difficult to treat exactly, even
in the test-fluid approximation. Severe simplifications have 
thus been imposed on theoretical models, ranging from 
presumed symmetries and boundary conditions to various
assumptions about
microphysics of the medium. Simplifying assumptions are
the more unavoidable when the disc gravity (playing a role of the 
external source to the gravitational field) 
can {\em not\/} be neglected and when one intends to
proceed analytically. Further restrictions are
dictated by applicability of the results to real cosmic objects.

\section{Stationary axisymmetric fields in general relativity}
\label{sec:relativity}          
There are at least three aspects in which the system of a rotating black
hole with an external source attracts relativists. First, due to the
non-linearity of Einstein's equations, the field of a multi-body system
is a traditional challenge where, in most cases, one does not manage
with a simple superposition. On the first post-Newtonian level, the
``celestial mechanics'' of gravitationally interacting bodies can be
kept linear (see Damour {\etal} \cite{DamSX91,DamSX92}),  but in the
strong-field regime the interaction may bring surprising features  that
have only been described in very few cases yet (for a two-body problem, 
the current ``state of art'' is the third post-Newtonian approximation
\cite{deABF01}).

The second point concerns manifestation of frame dragging due to the rotation
of the sources. Contrary to the Newtonian treatment (that does not
discriminate directly between static and stationary situations), the 
relativistic field is determined not only by mass-energy configuration, 
but also by its motion within the bodies: the inertial space can loosely be 
imagined as a viscous fluid mixed by the sources. In today's
``gravito-electromagnetic'' language: gravity has not only an electric
component, generated by the mass, but also a magnetic one, generated by
mass currents \cite{ThoPM86}. Let us only refer to somewhat casual selection
\cite{Pen83,Emb84,Mas85,JanCB92,CiuW95,LynKB95,KinP01} for further details at this point.

Needless to say, the third point is the presence of a {\bh} itself. 
Near such an extreme body, the deviations from Newtonian theory become dominant; 
in particular, the above mentioned implications of non-linearity 
and rotation reveal themselves prominently.

The class of stationary axisymmetric solutions of Einstein equations
is the appropriate framework for the attempts to include the gravitational
effect of an ``external'' source in an exact analytical manner. 
At the same time, such spacetimes are of obvious
astrophysical importance, as they describe the exterior of a 
body in equilibrium \cite{Isl85} (see
\cite{Lev68,Car73,Bar73b,Wil74,ButI76,Chan78,Emb84,ChinG93} 
for thorough expositions on stationary and axially symmetric spacetimes 
in general relativity).

In the Weyl--Lewis--Papapetrou coordinates ($t,{R},\phi,z$) of the
cylindrical type, the stationary axisymmetric metric can be
written as\footnote{Geometrized units are used in which $c=G=1$; the signature of
 the metric tensor $g_{\mu\nu}$ is ($-$+++). Greek indices run
 from 0 to 3 and Latin indices ($i$, $j$, \ldots) run from 1 to 3;
 indices from the beginning of the Latin alphabet ($a$, $b$,
 \ldots) will represent cyclic coordinates $t$ and $\phi$.
 Partial differentiation is denoted by $\partial$ or by a
 subscript comma, covariant derivative is denoted by $\nabla$.}
\begin{equation} \label{WLP}
 {\rd}s^{2}=-e^{2\nu}{\rd}t^{2}
  +{R}^{2}B^{2}e^{-2\nu}
  ({\rd}\phi-\omega{\rd}t)^{2}
  +e^{2\lambda-2\nu}({\rd}{R}^{2}+{\rd}z^{2}),
\end{equation}
where the unknown functions $\nu$, $B$, $\omega$ and $\lambda$
only depend on ${R}$ and $z$; $\omega$ is interpreted as the
angular velocity of inertial-frame dragging with respect to
observers at rest at spatial infinity. The metric coefficients
$g_{ab}$ have invariant meaning, they can be expressed in terms
of Killing fields
$k^{\mu}=\partial x^{\mu}/\partial t$ and
$m^{\mu}=\partial x^{\mu}/\partial\phi$:
\begin{equation}
 g_{tt}=-e^{2\nu}+\omega^{2}g_{\phi\phi}=k_{\iota}k^{\iota},
\end{equation}
\begin{equation}
 g_{t\phi}=-\omega g_{\phi\phi}=k_{\iota}m^{\iota},
\end{equation}
\begin{equation}
 g_{\phi\phi}={R}^{2}B^{2}e^{-2\nu}=m_{\iota}m^{\iota}.
\end{equation}
The corresponding sub-determinant is
$\det(g_{ab})=-{R}^{2}B^{2}$, so
\begin{equation} \label{WLPinverse}
 g^{tt}=-e^{-2\nu}, \quad
 g^{t\phi}=-e^{-2\nu}\omega, \quad
 g^{\phi\phi}=-e^{-2\nu}\omega^{2}+{R}^{-2}B^{-2}e^{2\nu}.
\end{equation}
The Einstein equations read
\begin{equation} \label{eqB}
 \vec{\nabla}\cdot({R}\vec{\nabla} B)=
 8\pi{R} B\left(T_{{R}{R}}+T_{zz}\right),
\end{equation}
\begin{equation} \label{eqomega}
 \vec{\nabla}\cdot\left({R}^{2}B^{3}e^{-4\nu}\vec{\nabla}\omega\right)=
 -16\pi Be^{2\lambda-2\nu}\,T^{t}_{\phi},
\end{equation}
\begin{equation} \label{eqnu}
 \vec{\nabla}\cdot(B\vec{\nabla}\nu)=
 \textstyle{\frac{1}{2}}{R}^{2}B^{3}e^{-4\nu}
 \left(\vec{\nabla}\omega\right)^{2}+
 4\pi Be^{2\lambda}\left(2T^{tt}+e^{-2\nu}\,T^{\sigma}_{\sigma}\right),
\end{equation}
where $\vec{\nabla}$ and $\vec{\nabla}\cdot$ stand for the gradient and
divergence in a flat three-dimensional space with coordinates
(${R}$,$\phi$,$z$); thus
$\vec{\nabla}X=(X_{,{R}},0,X_{,z})$ and
$\vec{\nabla}\cdot\vec{X}=
 {R}^{-1}[({R} X^{{R}})_{,{R}}+({R} X^{z})_{,z}]$
in the axially symmetric case.
Once $B$, $\omega$ and $\nu$ are known, $\lambda$ can be
integrated from equations
\begin{equation}
\eqalign{{\fl}B\lambda_{,z}-B_{,z}
 +{R}\,\Big(B_{,{R}}\lambda_{,z}+B_{,z}\lambda_{,{R}}-
  B_{,{R} z}-2B\nu_{,{R}}\nu_{,z}\Big) \\
 \label{lambda1}
 +\textstyle{\frac{1}{2}}{R}^{3}B^{3}e^{-4\nu}\omega_{,{R}}\omega_{,z}
 =8\pi{R} BT_{{R} z},}
\end{equation}
\begin{equation}
\eqalign{{\fl}2B\lambda_{,{R}}-2B_{,{R}}
 +{R}\,\Big[2B_{,{R}}\lambda_{,{R}}-2B_{,z}\lambda_{,z}-
  B_{,{R}{R}}+B_{,zz}-2B\left(\nu_{,{R}}^{2}-\nu_{,z}^{2}\right)\Big] \\
 \label{lambda2}
 +\textstyle{\frac{1}{2}}{R}^{3}B^{3}e^{-4\nu}
 \left(\omega_{,{R}}^{2}-\omega_{,z}^{2}\right)=
 8\pi{R} B\left(T_{{R}{R}}-T_{zz}\right).}
\end{equation}
The unknown metric functions are subject to boundary conditions on the
horizon (if there is one), on the symmetry axis and at spatial infinity.
The metric must be regular on the horizon and on the axis, and in the case
of an isolated source one also requires asymptotic flatness (the conditions
are discussed in \cite{Car73,Bar73b}, for example).

Outside of the sources, where $T_{\mu\nu}=0$, the simplest solution of 
equation (\ref{eqB}) with a satisfactory asymptotic behaviour (equation 
(\ref{B,asymp}) below) is $B=1$. The two remaining field
equations (\ref{eqomega})--(\ref{eqnu}) reduce to
\begin{equation} \label{eqomega,vac}
 \vec{\nabla}\cdot({R}^{2}e^{-4\nu}\vec{\nabla}\omega)=0,
\end{equation}
\begin{equation} \label{eqnu,vac}
 \vec{\nabla}^{2}\nu=
 \textstyle{\frac{1}{2}}{R}^{2}e^{-4\nu}(\vec{\nabla}\omega)^{2}
\end{equation}
and the relations (\ref{lambda1})--(\ref{lambda2}) yield
\begin{equation} \label{lambda,rho,vac}
 \lambda_{,{R}}=
 {R}\left(\nu_{,{R}}^{2}-\nu_{,z}^{2}\right)-
 \textstyle{\frac{1}{4}}{R}^{3}e^{-4\nu}
 (\omega_{,{R}}^{2}-\omega_{,z}^{2}),
\end{equation}
\begin{equation} \label{lambda,z,vac}
 \lambda_{,z}=
 2{R}\,\nu_{,{R}}\,\nu_{,z}-
 \textstyle{\frac{1}{2}}{R}^{3}e^{-4\nu}\,\omega_{,{R}}\,\omega_{,z}.
\end{equation}

Equations (\ref{eqomega,vac})--(\ref{eqnu,vac}) are often presented
in the form of the Ernst equation \cite{Ern68}
\begin{equation}
 -g_{tt}\vec{\nabla}^{2}{\mathcal E}=(\vec{\nabla}{\mathcal E})^{2}
\end{equation}
for a complex (Ernst) potential
${\mathcal E}\equiv -g_{tt}+{\rm i}\psi$
whose imaginary part $\psi$ is given by
\begin{equation}
 {R}\,\psi_{,z}=g_{tt}\,g_{t\phi,{R}}-g_{tt,{R}}\,g_{t\phi},
 \;\;\;\;\;
 {R}\,\psi_{,{R}}=g_{tt,z}\,g_{t\phi}-g_{tt}\,g_{t\phi,z}.
\end{equation}
$\lambda$ is found by a line integration as above.

\subsection{Essential features of the stationary axisymmetric
  spacetimes}
\label{features}

A black-hole horizon is a null hypersurface below which the 
spacetime is dynamical (see e.g.\ \cite{Car73,Car79,FroN98} for thorough accounts).
In Weyl-Lewis-Papapetrou coordinates, the horizon is located where
$\det(g_{ab})=-{R}^{2}B^{2}$ vanishes. In the vacuum case we can put $B=1$, 
so the horizon lies on the axis then, ${R}=0$. The {\bh} interior thus has 
to be studied in different coordinates, e.g.\ in spheroidal coordinates of the
Boyer--Lindquist type ($t,r,\theta,\phi$), introduced by the
transformation
\begin{equation} \label{BLcoord}
 {R}=\sqrt{\Delta}\,\sin\theta, \;\;\;\;\;
 z=(r-\mbh)\,y,
\end{equation}
where $\Delta=(r-\mbh)^{2}-k^{2}$ and $y=\cos\theta$, $\mbh$ being a scale
parameter (it represents the black-hole mass) and $(\pm)k$
determining where the horizon reaches up along the $z$ axis. The horizon
is now given by $\Delta=0$, i.e.\ $r=\mbh+k\equiv\rh$ (hence, $z=ky$; there
exist more black-hole horizons in general, but we mean the outermost one
here).

Important parameters of the horizon are its surface area $A$, 
surface gravity $\kappa_{\rmhor}$
and the angular velocity $\omega_{\rmhor}$ relative to infinity,
\begin{equation}
 A=2\pi\int\limits_{0}^{\pi}
 \sqrt{(g_{\theta\theta}\,g_{\phi\phi})_{\rmhor}}\,
 {\rd}\theta, \quad
 \kappa_{\rmhor}=|\nabla e^{\nu}|_{\rmhor}, \quad
 \omega_{\rmhor}=\omega(r=\rh);
\end{equation}
$\kappa_{\rmhor}$ and $\omega_{\rmhor}$ are constant all over the horizon.

In order to learn a ``true shape'' of the {\bh}, not
distorted by coordinates, the horizon must be represented
properly as a two-dimensional surface in a three-dimensional
Euclidean space. One can e.g.\ follow \cite{Sma73}.
The two-dimensional metric is first rewritten as
\begin{equation}
 {\rd}s^{2}=
 \frac{A}{4\pi}\,\Big[h^{-1}(y)\,{\rd}y^{2}+h(y)\,{\rd}\phi^{2}\Big],
\end{equation}
where $h(y)=4{\pi}A^{-1}(g_{\phi\phi})_{\rmhor}$.
An isometric embedding of the two-surface $(y,\phi)$ in
$\mathbb{E}^3$ with coordinates $(X,Y,Z)$ is then given by
\begin{equation} \label{embedd}
\fl
 X=\frac{A}{4\pi}\sqrt{h}\,\cos\phi, \;\;\;\;\;
 Y=\frac{A}{4\pi}\sqrt{h}\,\sin\phi, \;\;\;\;\;
 Z=\frac{A}{4\pi}\int\limits_{0}^{y}
 \sqrt{\frac{1}{h}\left(1-\frac{1}{4}h_{,y}^{~2}\right)}
 \;{\rd}y.
\end{equation}
If the horizon's Gaussian curvature
\begin{equation} \label{Gauss}
 {\mathcal C}_{\rmhor}=-8\pi^{2}A^{-2}\,h_{,yy}
\end{equation}
is negative in some place, the global embedding is impossible.

On the symmetry axis (${R}=0$), the regularity condition requires
$e^{\lambda}=B$, and therefore the metric adopts the form
${\rd}s^{2}=-e^{2\nu}{\rd}t^{2}+B^{2}e^{-2\nu}{\rm{}d}z^{2}$ here. 
Simplification of the formulae can also be expected in the
equatorial plane ($z=0$) if the spacetime is reflectionally 
symmetric with respect to it. This is usually the case for astrophysically
motivated considerations.

In the static case $\omega=0$. 
If $T^{{R}}_{~{R}}+T^{z}_{~z}=0$ (which is fulfilled with zero pressure), 
the metric (\ref{WLP}) acquires the Weyl canonical form
\begin{equation} \label{Weylmetric}
 {\rd}s^{2}=-e^{2\nu}{\rd}t^{2}
  +{R}^{2}e^{-2\nu}{\rd}\phi^{2}
  +e^{2\lambda-2\nu}({\rd}{R}^{2}+{\rd}z^{2}).
\end{equation}
Equations (\ref{eqB})--(\ref{lambda2}) reduce themselves to Poisson's
equation
\begin{equation} \label{eqnu,Weyl}
 \vec{\nabla}^{2}\nu\equiv
 {R}^{-1}\nu_{,{R}}+\nu_{,{R}{R}}+\nu_{,zz}=
 4\pi e^{2\lambda-2\nu}(T^{\phi}_{\phi}-T^{t}_{t})
\end{equation}
for $\nu$, and to relations
\begin{equation} 
 \eqalign{\label{lambda12,Weyl}
 \lambda_{,z}-2{R}\,\nu_{,{R}}\,\nu_{,z}=8\pi{R}\,T_{{R} z}, \\
 \lambda_{,{R}}-{R}\left(\nu_{,{R}}^{2}-\nu_{,z}^{2}\right)=
 4\pi{R} \left(T_{{R}{R}}-T_{zz}\right)}
\end{equation}
for $\lambda$. The linearity of Laplace equation allows for simple
superposition of vacuum Weyl fields.

From the astrophysical point of view, a given analytical solution is 
problematic if it contains features which should not be present in a
realistic source, such as
(i)~physical singularities on or above the horizon, or (ii)~bad 
asymptotic behaviour. The existence of a singularity follows --- according to
its type \cite{EllS77,Cla93} --- e.g.\ from the divergence of
scalars constructed from the metric tensor or from its derivatives, or
from the divergence of physical (tetrad) components of the Riemann
tensor. The Kretschmann invariant
$R_{\mu\nu\rho\sigma}R^{\mu\nu\rho\sigma}$ is usually checked first. However,
the spacetimes are known which do contain singularities, although all of
their curvature invariants vanish. It is thus difficult to verify the
{\em{}non-}existence of singularities; the only proof of regularity at a
given point is to find a coordinate system in which the metric
is smooth enough locally.

Astrophysically relevant is the region of outer communications
outside the horizon, but a relativist is interested in the 
black hole interior as well, because even there it is possible to
perform physical observations. In the Weyl--Lewis--Papapetrou
coordinates the metric is not defined below the horizon (this would
correspond to imaginary radius ${R}$) and must be extended there.
One asks about how is the {\bh} interior (and the singularity in particular)
influenced by the presence of external sources. The answer is unknown
even for quite simple superpositions. 

Far away from the horizon, one inquires for the asymptotic
behaviour of the field. That of an isolated stationary source
falls to zero in a specific manner --- the spacetime is
said to be asymptotically flat (e.g.\ chapter~11 in R~Wald's book 
\cite{Wal84}, or section~3 in J~Ehler's Festschrift \cite{BeiS00}). 
In case of the metric (\ref{WLP}), it must hold, for $r\rightarrow\infty$,
\begin{equation} \label{nu,asymp}
 \nu=-\frac{{\rm total~mass}}{r}+{\cal{}O}(r^{-2}),
\end{equation}
\begin{equation} \label{B,asymp}
 B=1+{\cal{}O}(r^{-2}),
\end{equation}
\begin{equation} \label{omega,asymp}
 \omega=\frac{2\cdot({\rm total~angular~momentum})}{r^{3}}
  +{\cal{}O}(r^{-4}),
\end{equation}
\begin{equation} \label{lambda,asymp}
 \lambda={\cal{}O}(r^{-2}).
\end{equation}

Although a suitable combination of solenoidal motions can 
satisfy the assumption of stationarity and axial symmetry, it is
natural to assume that the elements of the source follow the simplest
type of motion --- that along spatially circular orbits
(${R}={\rm const}$, $z={\rm const}$) with steady azimuthal
angular velocity $\Omega={\rd}\phi/{\rd}t$. Such a motion
follows the symmetries, so an
observer on a circular orbit experiences time-independent field around.
In order that the corresponding
four-velocity $u^{\mu}=u^{t}(k^{\mu}+\Omega m^{\mu})$
points inside the light cone,
$(u^{t})^{-2}=-g_{tt}-2\Omega g_{t\phi}-\Omega^{2}g_{\phi\phi}
 =e^{2\nu}(1-\hat{v}^{2})$
must be positive (we denoted $\hat{v}={R} Be^{-2\nu}(\Omega-\omega)$ 
the linear speed with respect to the local zero-angular-momentum 
observer). The latter condition is fulfilled if $\Omega$ falls within the
range
\begin{equation}
 \Omega_{\rm min}\equiv
 \omega-\frac{e^{\nu}}{RB} 
 \leq\Omega\leq
 \omega+\frac{e^{\nu}}{RB}\equiv
 \Omega_{\rm max}.
\end{equation}
On the horizon (where $e^{\nu}=0$), the permitted range narrows
down to a unique value $\omega_{\rmhor}$. In the above-given equations,
the four-velocity covariant components
\begin{equation} \label{gamma,ell}
 \eqalign{
 \gamma&\equiv -u_{t}=
 -u^{t}(g_{tt}+\Omega g_{t\phi})=u^{t}e^{2\nu}+\omega\ell,\\
 \ell&\equiv u_{\phi}=u^{t}g_{\phi\phi}(\Omega-\omega)}
\end{equation}
stand for the specific energy and the specific azimuthal angular
momentum with respect to an observer at spatial infinity.

In case of the circular orbit, the four-acceleration
$a_{\mu}=u_{\mu;\nu}u^{\nu}$ can be written as
\begin{equation}
 a_{\mu}=-{\textstyle{\frac{1}{2}}}g_{\alpha\beta,\mu}\,u^{\alpha}\,u^{\beta}
 =-\frac{g_{tt,\mu}+2\Omega\, g_{t\phi,\mu}+
 \Omega^{2}g_{\phi\phi,\mu}}{2e^{2\nu}\left(1-\hat{v}^{2}\right)}.
\end{equation}
It has at most two non-zero components, $a_{{R}}$ and $a_{z}$;
in general, $a_{z}$ is zero in the equatorial plane and non-zero
elsewhere, while $a_{{R}}$ vanishes if
\begin{equation} \label{Omegapm}
 \Omega=\Omega_{\pm}=
 \frac{-g_{t\phi,{R}}
 \pm\sqrt{(g_{t\phi,{R}})^{2}-g_{tt,{R}}\,g_{\phi\phi,{R}}}}
 {g_{\phi\phi,{R}}} \, ,
\end{equation}
where the upper/lower sign corresponds to a
co-rotating/counter-rotating orbit (these terms may be somewhat
misleading in spacetimes with large angular momentum). Among
equatorial geodesics, three particular cases are important: the
photon, the marginally bound and the marginally stable circular orbits. The
photon orbits delimit the regions where the circular motion is time-like; they
are given by equalities $\Omega_{\pm}= \Omega_{\stackrel{\rm
max}{\scriptscriptstyle\rm min}}$. The marginally bound orbits demarcate the
regions where particles on circular orbits have lower energy than 
necessary for the existence at spatial infinity; the limiting case is
given by ${\gamma}(\Omega=\Omega_{\pm})=1$. The marginally stable orbits
also bound the regions where circular motion is stable with respect to small
perturbations acting within the orbital plane. Rayleigh's criterion of 
linear stability gives the location of the marginally stable orbits where
$[\ell(\Omega=\Omega_{\pm})]_{,{R}}=0$. These orbits play a key role in the
theory of accretion discs: the matter is swept away from unstable
sectors. In particular, the innermost marginally stable orbit should
represent the disc inner rim. However, perturbations
{\em{}perpendicular} to the orbital plane can also be 
important.\footnote{It has been discussed quite recently that 
(non-self-gravitating) accretion discs are
prone to warping due to extraneous irradiation \cite{Pri96}.
This effect is, naturally, modified when self-gravitation is taken
into account \cite{PapTL98}.}
For the above-mentioned exact 
solutions of Einstein's equations, the frequencies of vertical
oscillations were given explicitly \cite{SemZ00b}.

A peculiar feature of rotating fields are the dragging effects. 
In accordance with Mach's ideas of relativity of motion and inertia (and
his interpretation of Newton's bucket experiment in particular), Lense
\& Thirring showed, in the early times of general relativity, that
dragging is present in Einstein's theory. Machian inspiration remains
alive within contemporary relativity --- see, for example, the
last-decade references \cite{JanCB92,CiuW95,LynKB95,LynBK99,KinP01} or
the T\"ubingen conference \cite{BarP95} and references therein. The
effect has obvious analogy in electromagnetism, where (electric)
currents generate the magnetic component of the field. An extreme
implication of dragging is the occurrence of the ergosphere in the
vicinity of ultra-compact rotating objects. In this region $\Omega_{\rm
min}>0$, thus the stationary observer cannot remain static (i.e.\ at
rest relative to infinity), albeit he still withstands the radial
attraction. The static-limit surface, given by $\Omega_{\rm min}=0$ and
hence by $g_{tt}=0$, also represents the set of points from where
signals emitted by static observers ($\Omega=0$) reach infinity with an
infinite redshift. Several processes working in the ergosphere have been
suggested, by means of which the rotational energy of the black hole
could be extracted without violating the second law of {\bh} dynamics
\cite{Pun01}.

With the metric (\ref{WLP}), $g_{tt}=0$ corresponds to
${R}B\omega=e^{2\nu}$ which can only be satisfied by ${R}\geq 0$, so the
static limit really lies outside the horizon;\footnote{In coordinates
which also describe the {\bh} interior, {\em{}two\/} horizons and
{\em{}two\/} static limits are found in general. We mean the outer
horizon and the outer static limit everywhere.} in particular, ${R}=0$
implies $e^{2\nu}=0$, so the static limit touches the horizon at the
axis.

\subsection{Sources}
\label {sources}
In order to complete the relativistic solution, one must also describe
the interior of sources --- where $T_{\mu\nu}\neq 0$. It is difficult 
to find a realistic interior solution, especially if it has to match 
smoothly the vacuum exterior and if we restrict ourselves to analytical 
approaches. The problem can be substantially simplified by assuming that
the source is infinitely thin: the solution is then vacuum-type
everywhere, with the energy-momentum tensor
$T_{\mu\nu}=g_{zz}^{-1/2}S_{\mu\nu}\delta(z)$ found from the 
discontinuity of the
normal field across the source as in electrodynamics (the appropriate
covariant method is known as the Israel's formalism \cite{Isr66,BarI91}).

The case of a thin equatorial layer (a disc) in a stationary
axisymmetric spacetime has notably been studied by
\cite{Led98,GonL00}. The disc lying on $z={\rm const}$ has zero
radial pressure inside ($S_{{R}{R}}=0$). At the same time, the 
vertical pressure
vanishes ($S_{zz}=0$) if the disc mid-plane coincides with 
the equatorial plane ($z={\rm const}=0$) of the reflectionally symmetric 
spacetime. In the Weyl--Lewis--Papapetrou coordinates, the non-zero 
$S_{\mu\nu}$ components read \cite{Led98}
\begin{equation} \label{Sab}
 S_{ab}=-\frac{\sqrt{g_{{R}{R}}}}{8\pi}
  \left(\frac{g_{ab}}{g_{{R}{R}}}\right)_{\!\!,z}.
\end{equation}
Using (\ref{WLPinverse}) and (\ref{lambda,z,vac}), this yields
\begin{equation}
 S^{t}_{t}=
 -\frac{e^{\nu-\lambda}}{8\pi}
 \Big[4\nu_{,z}(1-{R}\nu_{,{R}})-
 {R}^{2}e^{-4\nu}\omega_{,z}(\omega-{R}\omega_{,{R}})\Big],
\end{equation}
\begin{equation}
 S^{t}_{\phi}=
 -\frac{e^{\nu-\lambda}}{8\pi}{R}^{2}e^{-4\nu}\omega_{,z},
\end{equation}
\begin{equation}
 S^{\phi}_{t}=
 -\frac{e^{\nu-\lambda}}{8\pi}
 \left[4\omega\nu_{,z}-(1+{R}^{2}e^{-4\nu}\omega^{2})\omega_{,z}
 \right],
\end{equation}
\begin{equation}
 S^{\phi}_{\phi}=
 \frac{1}{8\pi}\,R\,e^{\nu-\lambda}\,
 \Big[4\nu_{,{R}}\nu_{,z}-
 {R} e^{-4\nu}\omega_{,z}(\omega+{R}\omega_{,{R}})
 \Big].
\end{equation}
The total mass and angular momentum of the disc are fixed by the
Komar integrals that, respectively, lead to
\begin{equation} \label{Md}
 {\md}=\frac{1}{2}\int\limits_{b}^{\infty}
 g^{tc}g_{tc,z}{R}\,{\rd}{R}
 =-\frac{1}{4}\int\limits_{b}^{\infty}
 \frac{g_{\phi\phi}^{2}}{{R}}
 \left(\frac{g_{tt}}{g_{\phi\phi}}\right)_{\!\!,z}{\rd}{R},
\end{equation}
\begin{equation} \label{Jd}
 {\jd}
 =\frac{1}{4}\int\limits_{b}^{\infty}
 g^{tc}g_{c\phi,z}{R}\,{\rd}{R}
 =\frac{1}{4}\int\limits_{b}^{\infty}
 \frac{g_{\phi\phi}^{2}}{{R}}\,\omega_{,z}\,{\rd}{R};
\end{equation}
${R}=b>0$ is the position of the disc inner rim. All the
$z$-derivatives are understood to be calculated in the limit
$z\rightarrow 0^{+}$.

If a {\bh} is present in the disc's centre, its 
contribution to the total mass and angular momentum need
to be included. For a given type of spacetime, the aggregate values
of the parameters can be written as sums, ${\mbh}+{\md}$ and
${\jbh}+{\jd}$ (cf.\ \cite{Car79}, chapter 6.6.1). The black-hole 
parts can be calculated as Komar integrals, or they can be inferred
by subtracting ${\md}$, ${\jd}$ from the totals, derived from the 
metric asymptotes (\ref{nu,asymp},\ref{omega,asymp}).
${\mbh}$ and ${\jbh}$ are related through Smarr's 
formula,\footnote{See Carter \cite{Car79}, equation (6.297), for
derivation and discussion. For the original formulation in Kerr
spacetime, see \cite{smarr73}.}
\begin{equation} \label{MH}
 {\mbh}=2\omega_{\rmhor}{\jbh}+\frac{1}{4\pi}\,\kappa_{\rmhor}A=
  2\omega_{\rmhor}{\jbh}+k\,.
\end{equation}
Further, if the term
\begin{equation}
D\equiv\left(S^\phi_\phi-S^t_t\right)^2+4S^t_\phi S^\phi_t
 =\frac{e^{2\nu-2\lambda}}{16\pi^{2}}\,
 \left(4\nu_{,z}^{~2}-{R}^{2}e^{-4\nu}\omega_{,z}^{~2}\right)
\end{equation}
is non-negative, then a stationary observer exists
with respect to whom the energy-momentum tensor assumes a
diagonal (``isotropic'') form
\begin{equation}
S^{\mu\nu}=\hat{w}u_{\rm iso}^\mu u_{\rm iso}^\nu+
\hat{P}v_{\rm iso}^\mu v_{\rm iso}^\nu,
\end{equation}
where $u_{\rm iso}^\mu=u_{\rm iso}^t(1,0,\Omega_{\rm iso},0)$
and $v_{\rm iso}^\mu=-{R}^{-1}(\ell_{\rm iso},0,\gamma_{\rm iso},0)$
are, respectively, observer's four-velocity and the unit basis vector in the
$\phi$-direction. Angular velocity of this observer is
\begin{equation}
 \fl
 \Omega_{\rm iso}
 =\frac{1}{2S^t_\phi}\left(S^\phi_\phi-S^t_t-\sqrt{D}\right)
 =\omega-\frac{2e^{4\nu}\nu_{,z}}{{R}^{2}\omega_{,z}}+
 \sqrt{\left(\frac{2e^{4\nu}\nu_{,z}}{{R}^{2}\omega_{,z}}
 \right)^{2}-\frac{e^{4\nu}}{{R}^{2}}}\;;
\end{equation}
$\gamma_{\rm iso}$ and $\ell_{\rm iso}$ denote
the corresponding specific energy and angular momentum at 
infinity.\footnote{The case of $D<0$ represents discs with
non-zero heat flow. Then the energy-momentum tensor must be
written in a more general form,
$S^{\mu\nu}=\hat{w}u_{\rm iso}^\mu u_{\rm iso}^\nu+
  \hat{P}v_{\rm iso}^\mu v_{\rm iso}^\nu+
  \hat{K}(u_{\rm iso}^\mu v_{\rm iso}^\nu+
  v_{\rm iso}^\mu u_{\rm iso}^\nu)$,
with $\hat{K}=\sqrt{-D/2}$ and
$\Omega_{\rm iso}=(2S^t_\phi)^{-1}(S^\phi_\phi-S^t_t)
 =\omega-2R^{-2}\omega_{,z}^{-1}e^{4\nu}\nu_{,z}$.
See \cite{GonL00}.}
The observer measures the surface density
\begin{equation}
 \hat{w}=
 \frac{\gamma_{\rm iso}^{2}S^{tt}-
 \ell_{\rm iso}^{2}S^{\phi\phi}}
 {u_{\rm iso}^{t}
 (\gamma_{\rm iso}+\Omega_{\rm iso}\ell_{\rm iso})}
\end{equation}
and tangential pressure
\begin{equation}
 \hat{P}=\frac{{R}^{2}u_{\rm iso}^{t}
 (S^{\phi\phi}-\Omega_{\rm iso}^{2}S^{tt})}
 {\gamma_{\rm iso}+\Omega_{\rm iso}\ell_{\rm iso}}.
\end{equation}
If $\hat{w}\geq\hat{P}>0$, the energy-momentum tensor can
represent two equal streams of particles moving on (accelerated)
circular orbits in opposite directions at the same speed
$\sqrt{\hat{P}/\hat{w}}$. This ``speed of sound'' was shown
\cite{GonL00} to be equal to the geometric mean of the local
prograde and retrograde circular geodesic speeds $\hat{v}_{\pm}$
as measured by the observer $u_{\rm iso}^{\mu}$:
\begin{equation}
 |\hat{v}_{+}\hat{v}_{-}|=\hat{P}/\hat{w},
\end{equation}
where
\begin{equation}
 \hat{v}_{\pm}=
 \frac{(v_{\rm iso})_t+(v_{\rm iso})_\phi\Omega_{\pm}}
 {(u_{\rm iso})_t+(u_{\rm iso})_\phi\Omega_{\pm}}\;;
\end{equation}
$\Omega_{\pm}$ are the angular velocities of the prograde and
retrograde circular geodesics (\ref{Omegapm}). Although
counter-rotating relativistic streams are hardly applicable to any
realistic astrophysical situation, they can be very illuminating
especially in the discussion of physical sources of the Kerr 
metric and other spacetimes.\footnote{Bi\v{c}\'ak~\& Ledvinka 
\cite{BicL93} have constructed pressure-less counter-rotating 
discs in the Kerr geometry,
employing Israel's technique \cite{Isr66,BarI91}.
Analogous approach was used by other authors  
in a different context of exact solutions; see e.g.\
Abramowicz, Arkuszewski \& Muchotrzeb \cite{AbrAM76}, 
Turakulov \cite{Tur84}, and Pichon \& Lynden-Bell \cite{PichL96},
or a particularly lucid paper by Bi\v{c}\'ak, Lynden-Bell \& 
Katz \cite{BicLK93}.
These works may appear rather academic for people who are
interested mainly in astrophysical applications of gravitating discs. 
However, a very pertinent question is addressed in these papers,
namely, under what conditions a material disc could represent
a source for the Kerr metric. 
Very recently, the ``displace, cut and reflect'' method has been extended 
to generate static thin discs with halos \cite{VogL03} and 
static {\em thick} discs \cite{GonL04}.}

Further physical requirements can be raised, namely, the weak energy
condition ($\hat{w}\geq 0$, $\hat{P}\geq -\hat{w}$), the dominant energy
condition ($\hat{w}\geq 0$, $|\hat{P}|\leq\hat{w}$) and the
non-negativity of pressure ($\hat{P}\geq 0$). Their combination gives
$\hat{w}\geq\hat{P}\geq 0$. The requirements are very simple in the
static case: if $\nu_{,z}(z=0^{+})>0$, then $\hat{P}\geq 0$ is equivalent 
to $\nu_{,{R}}\geq 0$ which is satisfied below the Lagrangian point of
zero field (where $\nu_{,{R}}=0$ and $\hat{v}_{\pm}^{2}=0$). In regions
with tension ($\hat{P}<0$) the particles would be attracted more by the
outer parts of the disc than by the central {\bh}, thus no circular
geodesics exist (then hoop stresses have to be employed to interpret the
disc matter); this typically happens in discs with much matter on
large radii. The other condition, $\hat{w}\geq\hat{P}$, is equivalent to
$\hat{v}_{\pm}^{2}\leq 1$. To summarize, the above constraints are in
fact ensured by the obvious condition $0\leq\hat{v}_{\pm}^{2}\leq 1$
which can be written explicitly as $1\geq 1-{R}\nu_{,{R}}\geq 1/2$.

\subsection{Solving Einstein's equations for a black hole with matter around}
Relativistic spacetimes are being found in three ways: 
(i)~by numerical solution of Einstein's equations, 
(ii)~by perturbation of previously known spacetimes, and 
(iii)~by exact analytical solution of Einstein's equations. Let us mention
several results of the above approaches on stationary axisymmetric
spacetimes that describe rotating black holes with additional matter.

Teukolsky \cite{Teu98} lists today's computational resources
and the appropriate, hyperbolic formulation of the
field equations as the main ``reasons why we are on the verge of
important advances in the computer solution of Einstein's
equations''. 
Whereas analytical prospects are restricted, the
present-day computers can handle almost any
situation. However, covering a sufficiently representative time-span
is still a problem. Furthermore, it is often impossible to recognise 
whether a given numerical solution represents a
typical situation or just a marginal case. Due to this lack of generality,
it is difficult to discern, analyse and interpret different
classes of solutions across boundless ranges of possibilities.
Nevertheless, numerical solutions provide explicit {\em examples}
of spacetimes and processes that might otherwise remain only
conjectural \cite{Leh01}.

Numerical spacetimes with a rotating {\bh} surrounded by a
stationary axisymmetric source were already constructed by
Lanza
\cite{Lan92} (a hole with a thin finite equatorial disc) and by
Nishida \& Eriguchi
\cite{NisE94} (a hole with a thick toroid); recently, a new
(``multidomain spectral'') method has been utilized to spheroidal
as well as toroidal fluid configurations \cite{AnsKM03},
with the possibility of inclusion of a central object in future. 
To mention just one
particular point, contrary to a common experience (also gained by
\cite{NisE94}) that the horizons inflate towards the external sources,
\cite{Lan92} ended with a prolate horizon (stretched along the
rotation axis) in certain cases (when the disc was strongly
counter-rotating with respect to the hole). It would be an interesting
consequence of the interplay between dragging from the hole and from the
disc if it were confirmed that this can indeed happen, even though under
extreme circumstances only.

A great deal of literature and several formalisms have been devoted to 
perturbations of {\bh} spacetimes.
Will \cite{Wil74} gave an explicit result for the
perturbation of the Schwarzschild metric by a (rotating) axisymmetric
weakly gravitating thin equatorial ring. This direct approach is however
not practicable for a rotating hole and/or for an extended
source with pressure (not mentioning more complicated situations).

The rotating case (specifically, the algebraically special vacuum case)
was discussed notably by Teukolsky \cite{Teu73} who succeeded in separating
the decoupled equations for the first order perturbations of a Kerr
{\bh} into the second-order ordinary differential equations. By
solving the ``gravitational'' Teukolsky equation, the perturbative
deformation of the Kerr horizon was calculated by \cite{Dem76}. The
stationary axisymmetric Green's function of the Teukolsky equation was
provided by \cite{Lin77}. The most comprehensive expositions of the
first-order {\bh} perturbation theory were given by Chandrasekhar
\cite{Chan79,Chan83} in the Newman--Penrose formalism.

In a related approach to the first-order perturbations of
rotating fields \cite{Chrz75,KegC79,Tor90,TorS99}, the perturbation
components are expressed in terms of a single (Debye) potential that
obeys a wave-like equation (adjoint of the Teukolsky master equation).
The problem of perturbation of the Kerr horizon was treated this way by
Chrzanowski \cite{Chrz76}. 
Yet another approach, not restricted to algebraically 
special backgrounds, was advanced in \cite{SarHB01}. 
Note that all these perturbative methods deal
with curvature components, as opposed to the earlier metric 
formalisms of Regge \& Wheeler, Zerilli and Moncrief.

In the charged case, the perturbation problem introduces coupling
between gravitational and electromagnetic quantities.
Conversion between gravitational and electromagnetic waves takes place. 
In the case of Reissner--Nordstr\"om {\bh}, these interacting
perturbations were studied by Bi\v{c}\'ak
\cite{Bic79} and, more recently, by Torres del Castillo {\etal}
\cite{TorC96}. A gauge invariant derivation of the basic equations
was given by Fernandes \& Lun in Schwarzschild space-time \cite{FerL96}, while
\cite{FerL97} discussed a rotating analogy and provided its link 
with the Teukolsky equation.

The second-order perturbation theory has also been under development
since the 1970s. For Schwarzschild black holes, it is surveyed in
Gleiser {\etal} \cite{GleNPP00}, while the rotating case is tackled in 
Campanelli \& Lousto \cite{CamL99}.

It should be remarked that solutions describing stationary sources
around black holes are not the primary aim of the 
perturbation strategy. Historically, attention has rather been devoted
to the stability of black-hole solutions, to the relaxation of black
holes into a stationary no-hair state, to non-stationary processes
of astrophysical significance (such as the behaviour of weakly
gravitating particles or waves in {\bh} backgrounds) and to
gravitational waves (see Thorne \cite{Tho98} for a survey).

One can add that the approximate (though non-perturbative) method for
superposing two (rotating) black holes was proposed by 
Krivan \& Price \cite{KriP98} to
fix initial data for both analytic and numerical computations of {\bh} 
collisions.

A perturbative approach is adequate in situations where the outside
source has only a very small effect on the field of the main body.
Whenever this is not the case, the superpositions have to be described
by solutions of the full, non-linear theory. At the end of Bi\v{c}\'ak's
survey \cite{Bic00}, one is encouraged ``not to cease in embarking upon
journeys for finding them, and perhaps even more importantly, for
revealing new roles of solutions already known'', because ``Is there
another so explicit way of how to learn more about the rich
possibilities embodied in Einstein's field equations?'' More recent
reviews of vacuum fields can be found in Bonnor {\etal}
\cite{Bon92,BonGM94}. For further details, see also the  canonical
monograph on exact solutions to Einstein's field equations by Stephani
{\etal} \cite{SteKMHH03}. 

Until now, no explicit exact solutions describing the system of a
rotating centre with an additional axisymmetric ring, disc or torus have
been found. Nevertheless, many stationary axisymmetric solutions of the
(electro-)vacuum Einstein equations are known that do generalise those
containing only isolated black holes. These were mostly obtained by
indirect methods known as ``generating techniques''. Two major
approaches, developed by the end of the 1970's --- the group-theoretic
techniques and the soliton-theoretic (or inverse-scattering) techniques,
work for spacetimes with two commuting symmetries. (See Kordas \cite{Kor99} 
for a review and \cite{Cos80,Cos81,Cos82,HoeD84,Mai00} for more detailed
analysis and interrelations between different formulations, e.g.\ those
of Kinnersley \& Chitre; Maison; Belinskii \& Zakharov; Harrison;
Hoenselaers, Kinnersley \& Xanthopoulos; Hauser \& Ernst; Neugebauer;
Kramer \& Neugebauer; or Alekseev.) Other related methods have been
proposed more recently, e.g.\ the simplification of the Hauser--Ernst
integral equation by Sibgatullin \cite{Sib84}, ``monodromy data
transform'' by Alekseev \cite{Ale01} (also \cite{KleR98}), static
gravitational multipoles \cite{GutM85} (cf. \cite{Que92}) and their
superposition with stationary fields \cite{GutM88} by Gutsunaev \&
Manko, linear transformation by Quevedo \cite{Que92} or 
``finite-gap'' (algebraic-geometric) solutions by Korotkin \& Matveev 
\cite{Kor97}. Most of them have been compared in reference
\cite{AleG96}. (For other approaches, see e.g.\
\cite{Tan79,Cle00} or the results of Nakamura and Kyriakopoulos,
referred to and worked out by \cite{Ter90}; cf. also \cite{Vei85}.)

The crucial point of the soliton generating techniques is a solution of
two linear differential equations (Lax pair) the integrability
conditions of which are exactly the Einstein equations (namely the
Ernst equation). The linear problem can be tackled in order to generate
new solutions from the known ones: given some (``seed'') metric with two
Killing vectors, it yields a different metric of the given type (the
procedure is often called B\"acklund transformation in analogy with the
technique used for the KdV equation). In such a manner, many known
spacetimes were reproduced, but also broad families of new solutions
were provided characterised by arbitrarily large sets of free
parameters. Only a very restricted number of them have been given a 
clear physical interpretation, however. Though several of these results
perhaps represent a rotating {\bh} in an ``external'' gravitational
field (e.g.\ \cite{QueM91,Man92,ManN92,BreGMD98,ChauD97a,ChauD97b}), none
of the latter has yet been firmly linked with a concrete
realistic body such as ring, disc or torus (but cf.\ section~{\sc{}iv} 
in Letelier \& Oliveira \cite{LetO87}). Namely, when performing the 
B\"acklund transformation, it is difficult to require specific physical
properties of the spacetime being generated --- one rather tries what
will come out. See \cite{Que92} for a hint of how to overcome this. More
recently, Letelier \cite{Let99} interpreted a solution obtained from a
Weyl seed by the inverse-scattering transformation with an even number
$N$ of solitons as a superposition of this Weyl solution --- i.e.\ of a
given set of static multipoles --- with $N/2$ Kerr(-NUT) solutions,
represented by $N/2$ rotating axial bars in Weyl-type coordinates.

As an example, let us give the attempt \cite{ZelS00} which was
``cultivated''  from a general Weyl metric by a real-two-soliton version
of the Belinskii-Zakharov method. The resulting metric was written as a
generalisation of the Kerr(--NUT) solution and many of its properties
were proven satisfactory. Choosing the inverted first Morgan--Morgan disc
(see below) as a seed, various characteristics of the superposition were
plotted \cite{Sem02c} against the rotation parameter of the central
hole, the mass of the disc and its inner radius. The characteristics
simplify very much in the limit when the {\bh} becomes extreme. One of
the interesting features of this limit is the vanishing of the
``external'' gravitational field on the horizon \cite{Sem02a}. This
effect of expulsion of the external (stationary axisymmetric) fields
from rotating (and/or charged) black holes, analogous to the Meissner
effect in (super)conductors, was observed in magnetic fields before (see
the recent survey \cite{BicL00}). However, a serious problem has been
found in the equatorial plane: the black-hole horizon has a sharp
narrowing there. This is caused by the appearance of a supporting
surface between the hole and the source outside it \cite{Sem02c}. 
Indeed, the possible problems with physical interpretation of the
solution (e.g.\ negative density or  pressure or matter moving at a
superluminal speed) also involve the occurrence of supporting singular
surfaces or lines (``struts'') that indicate that a given system of
bodies cannot remain in stationary (or static) equilibrium. 

The struts may equally well ``spoil'' a static superposition,
though this case almost reduces just to adding the two potentials,
$\nu=\nu_{1}+\nu_{2}$. The singularity may appear
in calculating the second metric function,
\begin{equation}
 \lambda=
 \int\limits_{\rm axis}^{{R},z}
 {R}\,\Big[\left(\nu_{,{R}}^{2}-\nu_{,z}^{2}\right)\,{\rd}{R}
  +2\nu_{,{R}}\,\nu_{,z}\,{\rd}z\Big].
\end{equation}
Consequently, even the {\em static} axisymmetric case has only been
afforded a few realistic superpositions. Let us refer to two of
them involving a Schwarzschild {\bh}: the superposition with
an infinite annular thin disc, obtained by Lemos \& Letelier
\cite{LemL94} by inversion of the first counter-rotating finite
disc of Morgan \& Morgan \cite{MorM69} and the one with an
inverted isochrone thin disc, obtained by Klein \cite{Kle97}. The
Morgan--Morgan family of static axisymmetric finite thin counter-rotating 
discs \cite{MorM69} is given by Newtonian surface density
\begin{equation} \label{density}
 w^{(m)}({R}\leq b)=
 \frac{(2m+1){\md}}{2\pi b^2}
 \left(1-\frac{{R}^{2}}{b^{2}}\right)^{\!m-1/2} \;\;\;\;
 (m=1,2,\dots).
\end{equation}
Since the zeroth member is clearly singular at the rim, most
attention was devoted to the first ($m=1$) member. It appeared
to be a prototype of a simple and physically
meaningful Weyl field and it was also considered as a limit case of a
more general class of {\em stationary} discs \cite{Ans01,FraK01}. 
The corresponding annular case, obtained by the inversion 
$R\rightarrow\frac{b^{2}R}{R^{2}+z^{2}}$, 
$z\rightarrow\frac{b^{2}z}{\rho^{2}+z^{2}}$, 
was shown to really allow for physically satisfactory situations 
\cite{LemL94,SemZZ99a,SemZZ99b,SemZ00a,ZacS02}.
There is nevertheless a problem at the rim of these discs, already 
suspected by checking the radial derivatives of (\ref{density}):
the $m$-th member of the family has infinite $m$-th and higher 
derivatives of the density at the outer rim. Calculation of the Kretschmann 
curvature invariant reveals that the rim is really singular and that the 
singularity is inherited by the inverted discs \cite{Sem01,ZacS02}.
The superpositions of a Schwarzschild {\bh} with (inverted)
``higher'' Morgan--Morgan counter-rotating discs were studied in
\cite{Sem03}. Comparison of the results obtained for the first
ten of them showed that the properties of the disc are very sensitive
to the local density profile. It will be desirable to also consider 
different classes of annular thin discs in superpositions, preferably 
of those not having any such drawback at the rim.

We have already explored one different family of 
annular disc solutions, namely, the case when density is a
power-law in Weyl radius,
\begin{equation}
  w^{(m,n)}({R}\geq b)=\frac{\prod\limits_{l=1}^{m}(ln+1)}{m!~n^m}
            \frac{{\md}b}{2\pi R^3}
            \left(1-\frac{b^n}{R^n}\right)^{\!m}.
\end{equation}
However, these discs also turned out not to be regular at the rim;
$(m+1)$-st radial derivative of their potential diverges, 
although all the radial derivatives of density are finite 
there, and the first $m-1$ derivatives even vanish. In a different
context of galactic stellar discs, an interesting approach
has been developed by Sibgatullin, Garcia \& Manko \cite{SibGM03}
who examined a new method of reconstruction of the surface
density distribution of a finite size disc with a central {\bh}.

Allowing for self-gravitation of the matter around the horizon, 
one has to reduce the generality of the problem elsewhere. 
Whereas test discs are treated within stationary axisymmetric 
settings, analytical solutions with exact inclusion of the ambient 
material have only been studied in more detail in static cases yet. 
This is of course a serious limitation, because the collapsed cores 
of galaxies as well as ultra-compact components in binaries are likely to 
rotate, perhaps very rapidly, which means that inertial dragging should be 
important there in addition to the effects of pure scalar attraction. 
Moreover, the composite spacetimes presented in the literature are constructed 
with discs whose field can be expressed in a tractable form rather than with 
those which really follow from some model of accretion. However, some 
superpositions --- such as e.g.\ those involving thin annular discs 
obtained by inversion of the Morgan--Morgan solutions      
\cite{LemL94,SemZZ99a,SemZZ99b,SemZ00a,SemZ00b,ZacS02}            
--- were shown to possess physically acceptable properties within             
an astrophysically plausible range of parameters. As such, they           
can indicate some of the effects that could occur as a                 
consequence of non-negligible gravitational contribution of the             
disc (see \cite{Sem02b} for a more thorough survey).      

Let us return to the linear problem (Lax pair of linear equations)
associated with the Ernst equation. In order to control the physical
properties of the solution being constructed, it is desirable to
translate the physical boundary conditions of the Ernst equation into
the language of quantities which appear in the linear problem and then
solve the latter (rather than to apply a ``random'' B\"acklund
transformation to some metric \cite{Let89}). This leads to
Riemann--Hilbert problems known from the theory of completely integrable
differential equations \cite{Neu96,KleR98,AnsKMN02}. Tackling the
stationary axisymmetric boundary value problem with the help of the
linear system, the field of two physically relevant types of sources
has been discussed: that of black holes and that of finite thin discs 
(\cite{Mei00,Neu00,Ans01,FraK01,NeuM03} and references therein). 
The hope that the above methods could be used to describe superpositions 
of both (e.g.\ \cite{Neu00,NeuM03}) has recently been supported by 
Klein \cite{Kle03} who presented an exact space-time of a regular 
black hole surrounded by an infinite annular disc as a subclass of 
Korotkin's solutions to the Ernst equation, obtained by Riemann-surface 
techniques \cite{KorM00}. The resulting metric is given in terms 
of theta-functions, being asymptotically flat under suitable choice of 
parameters. Klein's solution seems to involve cases with physically 
acceptable source disc.

\label{testfluid}
The limit case should be briefly mentioned when gravitation of the external
matter is ignored and the central rotating {\bh} fully determines
the field. Then the spacetime geometry is described by Kerr metric and the 
equilibrium configurations can be found analytically 
\cite{Bar73a,FisM76}, at least if the fluid obeys a simple
equation of state (such as a barytropic relation) and remains in 
dynamical equilibrium with purely rotational motion (four-velocity
${\vec{u}}=u^\phi\vec{e_\phi}+u^t\vec{e_t}$). This case is
over-simplified from the astrophysical point of view, but it
captures several crucial relativistic effects which do not 
occur in the Newtonian limit --- namely, the possibility of fluid 
overflow across the inner edge, which acts, effectively, as
the Lagrange L$_1$ point in a binary system.

The relativistic Euler equation can be integrated in terms of
specific enthalpy, $W\equiv-\int(\epsilon+P)^{-1}\rd{P}={\ln\,}|u_t|-F(l)$,
which can be further expressed as a function of $l(R,z)$, the
profile of angular momentum density
\cite{AmeLA89,AbrHG83,AbrJS78,KozJA78}. 
The only unknown function 
that needs to be specified is the rotation law, 
either in terms of $l=-u_\phi/u_t$ (which is a conserved quantity), 
or angular velocity $\Omega=u^\phi/u^t$. 
Rotation can be differential but the choice is not completely arbitrary 
because, according to Rayleigh's criterion for linearly stable 
configurations, $l$ must be increasing monotonically outwards from 
the rotation axis (a borderline case of constant $l$ was discussed
in detail). $W$ plays the role of an
effective potential for inertial forces and determines equilibrium
configurations in which the equipotential and isobaric surfaces
coincide, so this way constant pressure surfaces can be found. 
For example, in a special case of marginally stable configuration with
$l=\mbox{const}$, one finds $W(R,z)={\ln\,}|u_t|$.

\begin{figure*}[tb]        
\begin{center}  
\includegraphics[width=0.45\textwidth]{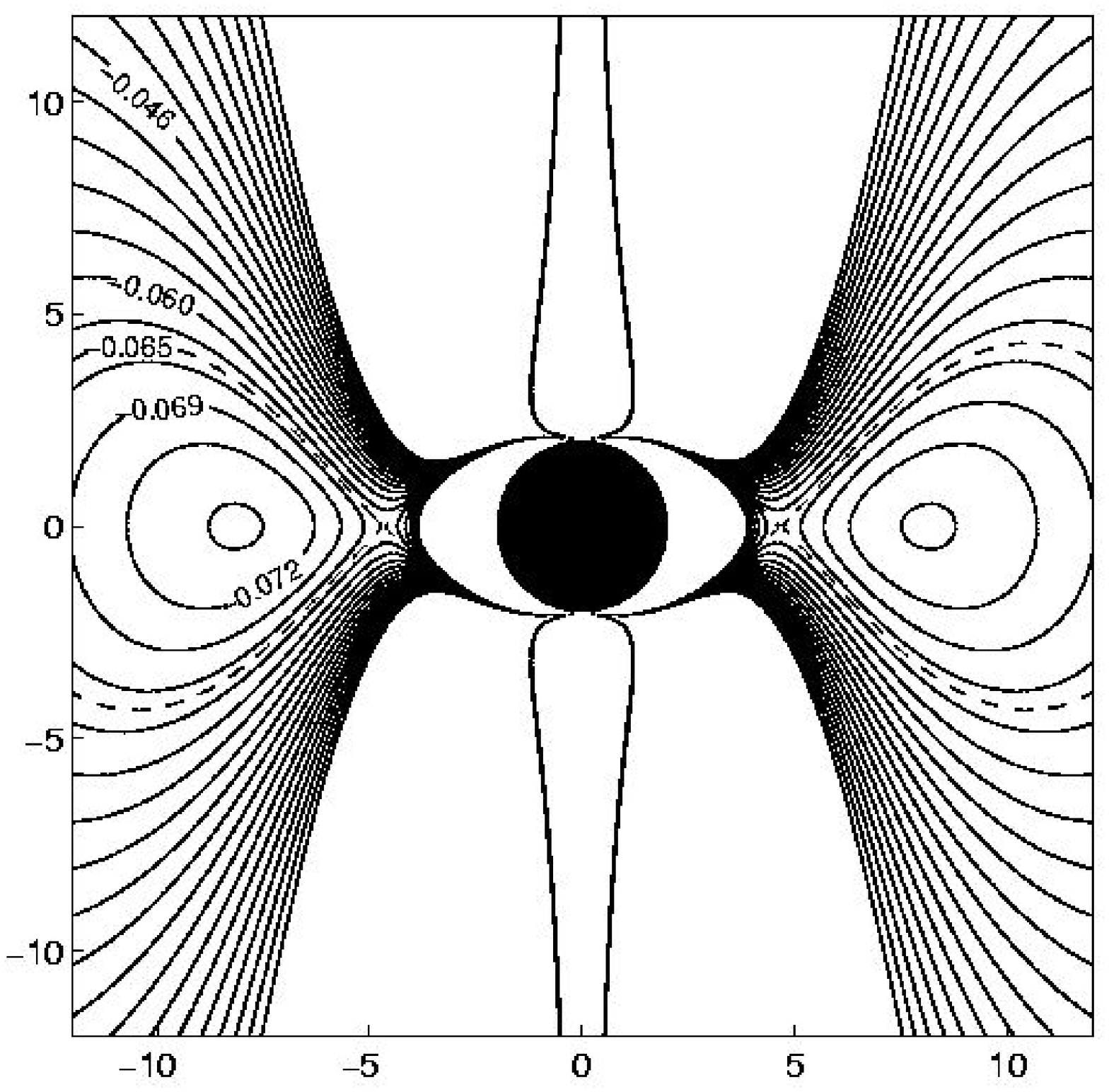}    
\hspace*{5mm}         
\includegraphics[width=0.45\textwidth]{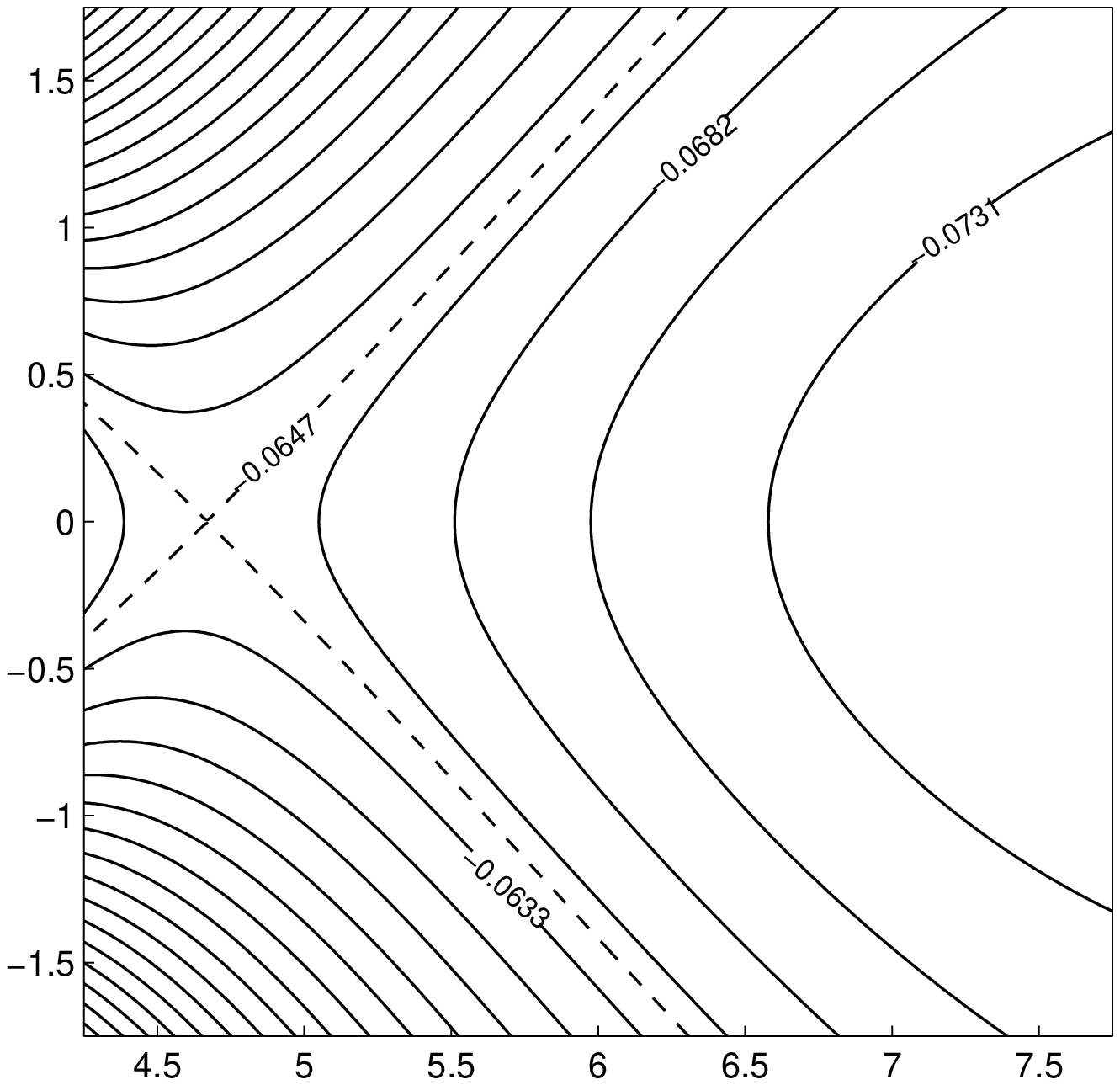}
\end{center}    
\caption{The equipotential surfaces of effective potential $W(x,z)$ are 
shown in a meridional section $(x,z)$ of the Schwarzschild metric.
Left: Contours of $W=\mbox{const}$ are plotted with solid lines.
The critical self-crossing contour, $W\doteq-0.065$, is
indicated with a dashed line. Equilibrium is possible if the
fluid (in pure rotational motion) 
fills one of the closed surfaces. Notice that these configurations
can extend far from equatorial plane. The black hole horizon is indicated 
by a circle of radius $\rs=2M$ around the origin. 
Right: A detail is shown of the structure of contours near the cusp. 
Fluid overflow provides the possibility of relativistic accretion 
across the cusp without the need for
viscosity. This effect also stands behind the origin of the runaway 
instability \cite{AbrJS78,KozJA78}.} 
\label{fig:w0}        
\end{figure*}

The fluid can establish
a stationary configuration provided that the $W$-isosurface
is closed (see figure~\ref{fig:w0}), otherwise overflow and subsequent 
accretion take place. Let us recall that equilibrium 
configurations with their elongated cross-section (bounded
by the critical, self-crossing lobe) {\em{}resemble\/} the shape of 
Roche lobes. The possibility of
fluid transport across the cusp occurs when the critical lobe is filled;
accretion can thus proceed without a necessary action of viscosity.
However, the two situations are of very different nature. In the former
case, it is the combined gravitational effect of
the binary together with (Newtonian) centrifugal force which define
the lobes, while, in the latter case, the only source of
gravity is the central body with relativistic effects playing a crucial
role.

\subsection{Gravitational effects of black-hole discs}
\label{gravity}
Compared with the vacuum case (with a {\bh} but 
without a disc), modifications of the
observable characteristics can be linked with the presence of
self-gravitating matter. 
Various approaches have been adopted in order to construct
solutions and discuss motion in the gravitational field of 
black-hole discs. One can start, e.g., by
constructing equilibrium structures of self-gravitating polytropic 
thick discs \cite{BodC92}. Formalism was 
developed also for perturbative solution of a
(slowly rotating, weakly gravitating, geometrically thin) ring 
around a Schwarzschild {\bh} \cite{Wil74,Wil75}. Furthermore, 
equatorial motion was explored in pseudo-Newtonian
studies \cite{KhaCh92} of a rotating {\bh} surrounded by a massive thin
equatorial ring, and in \gr\ analysis of the corresponding static case
\cite{Chak88}.\footnote{The 
pseudo-Newtonian approach was devised in order to introduce
effects of general relativity into accretion models \cite{PacW80}.
The black hole is treated as a Newtonian body whose gravitational
field is determined by Paczy\'nski--Wiita potential, 
$-\mbh/(r-\rs)$. This simple
expression mimics the gravitational field of a non-rotating
{\bh}; in particular, it reproduces the radii of marginally 
stable and marginally bound circular orbits of free test particles. 
Different variants of the pseudo-Newtonian potential were proposed in order
to include rotation of the {\bh} and to obtain better agreement with 
relativity (see \cite{SemK99} for references).
The pseudo-Newtonian potential does not aspire to represent any gravitational
theory, and the corresponding potential is not required to satisfy any 
field equations. This fact, however, does not diminish the practical 
value of the model, which applies even to regions
with strong gravity and captures qualitative features of motion near the
horizon.} The accretion-disc
self-induced field was also discussed in connection with the disc's vertical
structure and stability. It turns out that certain modes of instability
can be damped whereas others are amplified by the disc's gravity 
\cite{PapS91,AndTCh97}. Fully relativistic
results employ numerical spacetimes \cite{Lan92,NisE94}. 
Within these solutions, the computation of emission-line profiles 
were performed; the case of moderately massive thin finite discs 
was treated by \cite{KarLV95}, and that of
heavy toroids by \cite{UsuNE98}.\footnote{Let us mention here two
examples of the effects that we observed recently in studying the
properties of the Lemos--Letelier superposition. By generating a sequence of
superpositions parameterised with relative disc mass, and demanding that
(i)~all the disc matter be interpretable as two equal counter-rotating
streams of particles on stable time-like equatorial circular geodesics
and that (ii)~the inner disc rim be fixed at the least possible
radius, it was found \cite{SemZ00a} that heavier discs of the family can
reach considerably closer to the horizon not only in Schwarzschild $r$,
but also in terms of physical measures --- circumferential radius 
$r\exp(-\Psid)$ and proper radial distance from the horizon.}
Another issue is the oscillation of the disc material excited by a
small perturbation in the horizontal (${R}$ and $\phi$) or vertical
($z$) direction \cite{Kat01}. In the static case, the formulae for the
frequencies of free horizontal and vertical oscillations give
\cite{SemZ00b}\footnote{The perturbations are considered non-gravitating
here, i.e.\ the spacetime metric is not perturbed. Allowing the
perturbation to generate its own self-consistent field would only have a
second-order effect on the frequencies.}
\begin{equation} \label{kappa,Weyl}
 \kappa^{2}=\frac{e^{4\nu-2\lambda}}{1-{R}\nu_{,{R}}}\,
  \Big(\nu_{,{R}{R}}+4{R}\,\nu_{,{R}}^{3}-
  6\nu_{,{R}}^{2}+3R^{-1}\nu_{,{R}}\Big),
\end{equation}
\begin{equation} \label{Omegaperp}
 \Omega_{\perp}^{2}=
 \frac{e^{4\nu-2\lambda}}{1-{R}\,\nu_{,{R}}}
 \Big[\nu_{,zz}-4\nu_{,z}^{2}\left(1-2{R}\,\nu_{,{R}}\right)\Big].
\end{equation}
Without the external source, $\Omega_{\perp}^{2}$ reduces
to the square of the Schwarzschild orbital frequency,
$\Omega_{\pm}^{2}=M(M+\sqrt{{R}^{2}+M^{2}})^{-3}$, whereas $\kappa^{2}$
remains different due to the pericentre precession. It was found
that perturbations in the horizontal direction are more
important for the stability of light discs, whereas the vertical
perturbations are more dangerous for heavy discs: for 
superpositions of the given type, the curve of marginal vertical
stability is more restrictive than that of marginal horizontal
stability for ${\md}>0.2296M$. It even fully excludes discs with
${\md}\geq 2M/7$ \cite{ZacS02}. Disc self-gravity makes oscillations
of the inner parts faster and the region of the horizontal mode
trapping somewhat smaller.

When the black hole and the disc are allowed to rotate, 
manifestations of self-gravity 
could be even more pronounced, because in relativity the kinetic 
energy (as well as any other form of energy) also generates the field. 
Corresponding mass-energy currents give rise to frame
dragging. {\em{}Whereas the mass of the black hole is probably 
dominant in a real
accretion system, the disc can bear much of the angular momentum, thus
modifying the gravito-magnetic field of the central object significantly.} This
could be decisive for the mechanisms of rotational energy extraction. 
Note, however, also the possibility of having accreting systems where the 
disc/torus mass can actually be comparable to that of the central black 
hole and which can represent a transient state of
gravitational collapse. In this context, the most popular
system is a close binary of ultra-compact objects. Late evolutionary stages 
of such systems (in-spiral, merger and the ring-down phase) and
the fate of debris have been
studied, numerically and perturbatively, as a tentative source of 
gravitational waves (see \cite{Hug00} and references therein)
and also as an engine behind gamma-ray bursts
\cite{DaiM97,NarPP92,PruWH02,RosDTP00,RufJ99}.
In particular, at late stages of the neutron star--black hole binary
in-spiral, the neutron star could tidally break up
\cite{MasNE98,Mes01,NisLEA96,vPut01,Tho98}.

Two-dimensional toroidal shapes similar to the one shown in 
figure \ref{fig:w0} were originally 
invented mainly as a model for large and very diluted flows which
probably exist in quasars and similar objects. These considerations
have been continuously refined and discussed in various context
(recently, e.g.\ \cite{IguCA96} and references cited therein). 
As mentioned above, already in 
early works it was recognised that self-gravity imposes limits on this
scheme and should be taken into account in calculations.
Self-gravity is also relevant for a recently proposed
idea \cite{ZanRF03} that non-stationary tori
could be formed in accreting binaries. A relatively 
small size and steep pressure gradients of these configurations put them in 
a corner of parameter space where they may show high-frequency 
quasi-periodic oscillations \cite{vdK00}, which makes them very
attractive to astronomers, but formation of
such structures still needs to be addressed. 

\section{Astrophysical discs in Newtonian regime}
\label{sec:astrophysical}
From the point of view of gravitational theory, and also for the
nature of the accretion flow, the most important part is the inner region
near to the {\bh}. However, the gravitational field of actual discs has
substantial and even, possibly, dominant influence on their structure 
also at large distances (exceeding several $\times10^2\rs$;
in physical units,
$\rs=2G\mbh/c^2\dot{=}2.95\times10^{5}(\mbh/M_{\sun})\,$cm) where 
the self-gravity is
comparable to the central field. Newtonian physics is sufficient at
those distances. The
importance of the disc self-gravity is often checked by Toomre's
parameter, $Q={\kappa}c_{\rm{}s}/(\pi\Sigma)$: the local stability
criterion with respect to axially symmetric perturbations requires
$Q>1$ \cite{Too64,Too77} ($c_{\rm{}s}$ is a local sound speed).
Spiral structures may be produced already at $Q\sim2$--$4$. 
Depending on the radial distribution of the
thermodynamic quantities, global instabilities can indeed 
occur and play a key role in enhancing the accretion rate \cite{LinP96}.
(Of course, at this point one must recall the theories of galactic dynamics
\cite{BinT87}.) This may facilitate angular momentum transport and
provide a feeding mechanism for central black holes
\cite{FukHW00,HelS94,ShlBF90}.
Again, we concentrate on discs in galactic nuclei although similar 
mechanisms are relevant also for discs around stellar-mass black holes
and, to a certain degree, even for circumstellar discs in star-forming
regions.

\subsection{Effects of self-gravity derived from simple arguments}
In addition to the classical equations of mass, momentum and energy
conservation, the description of Newtonian 
gravitating fluids must include Poisson's equation,
\begin{equation}
\Delta\Psid(\vec{r})=4{\pi}\gconst\rho(\vec{r'}),
\label{eq:poisson}
\end{equation}
which links the mass density distribution ${\rho}$
in the disc and the self-gravitating potential $\Psid$ ($\vec{r'}$
refers to the source and $\vec{r}$ is the field point). To understand
the signatures of self-gravity expected in the fluid structure and
motion, one can consider a purely rotating, steady-state system with 
zero viscosity and no significant radial pressure gradient. 
The radial component of the
Euler equation (in cylindrical coordinates) is then:
$\Omega^2R=\nabla_{\!R}\,\Psih+\nabla_{\!R}\,\Psid$.
In the absence of rotation, steady solutions are forbidden and
the fluid falls towards the centre in a
short time scale. Free fall can be avoided thanks to the
centrifugal term which compensates the total 
gravitational attraction $g_R \equiv -
\nabla_{\!R} ( \Psid + \Psih)$. If radial self-gravity is
unimportant, that is if $\nabla_{\!R} \Psid \ll \nabla_{\!R} \Psih$,
then rotation of the fluid is determined by the
{\bh} mass only, $\Omega^2 \sim
\Omega^2_{\rm K}= \frac{1}{R}\nabla_{\!R} \Psih$. This assumption
underlies most disc models in astrophysics. In particular,
Newtonian geometrically thin disc models fall within this category
\cite{Pri81}. In the opposite situation, it is evident from 
the simple form of Euler's equation that, as the disc carries some mass, 
angular velocity is expected to increase in order to
prevent accretion. This implies that (radially)
self-gravitating discs are expected to rotate faster than
non-self-gravitating ones.

From the previous estimate one can conclude that heavy (fluid) discs
should exhibit a super-Keplerian rotation curve 
(in terms of $\Omega(R)>\Omega_{\mbox{\tiny{}K}}(R)$).\footnote{It was 
demonstrated \cite{Hur02} that the rotation curve of maser spots in 
NGC\,1068 could be explained by gravitational attraction of the 
disc around the central {\bh}.} Keplerian rotation
could still be maintained in a self-gravitating disc, but this would
be a surprising result of some fine tuning among various
quantities --- pressure gradients, advection of momentum, effects of
turbulent transport, etc. Neither can the possibility of sub-Keplerian 
rotation be ruled out. Such a case could occur in the outermost disc regions if
$\nabla_{\!R}\Psid<0$, for instance due to the presence of a dense region
near the edge or due to significant negative pressure gradients. 
Further, viscous stresses can switch the accretion.

Let us now analyse the vertical component of the Euler equation in the
same idealized situation as described above 
(again in the Newtonian limit):
$\rho^{-1}\nabla_{\!z}P=-\nabla_{\!z}\Psih-\nabla_{\!z}\Psid$. 
This is hydrostatic equilibrium in $z$-direction which
tells us that central gravity acts against pressure gradients and
tends to gather material around the equatorial plane of the disc.
For a bare central point mass, the magnitude of the vertical component 
of gravity decreases with radius. As an
inevitable consequence, the disc tends to `flare' (its thickness being
roughly proportional to the distance from the axis, or even steeper, and
it can wildly oscillate in the vertical direction). 
The atmosphere of the disc (i.e.\ its less dense region
away from the equatorial plane) gets thicker where the gravity is weaker. 
The geometrical semi-thickness of the disc then satisfies the relation
${\mathcal R}\equiv\frac{{\rd}\,\ln H}{{\rd}\,\ln R}>0$. 
Models show such flaring in the non-self-gravitating regime. 
Regarding the vertical direction, the disc
self-gravity is significant if $\nabla_{\!z}\Psid\sim\nabla_{\!z}\Psih$. 
For any fluid element
this results in enhancement of the vertical gravity; thus, a
vertically self-gravitating disc should be geometrically thinner and
more dense than a non-self-gravitating one. As will be shown below, 
some accretion disc models even yield ${\mathcal R}<0$ in those
regions, so in these circumstances the disc thickness cannot exceed a certain
limiting value. This is important for the disc irradiation (and
self-irradiation). 

Some fundamental properties of discs and tori can be understood in 
the apparently simple framework of Maclaurin spheroid
theory. Early analytical investigations by Poincar\'e, Dyson and, later,
by Chandrasekhar \cite{Chan69} have been supplemented with recent
numerical investigations (Hachisu \cite{Hac86,HacE83}). These 
works shed light on possible shapes and dynamical characteristics of
steady, self-gravitating polytropes in the inviscid limit. Both
axially and non-axially symmetrical equilibrium patterns have been
discussed. It turns out that equilibrium sequences can describe any 
conceivable system (depending on the polytropic index of the gas):
these include multiple star systems, tri-axial (Jacobi) ellipsoids,
some very strange shapes (`pear', `amonit', etc.), but also discs,
tori and rings. In the case of global solid-body rotation, discs, tori, 
and rings occupy the bottom-right part in the usual diagram of
rotation frequency versus angular momentum. Furthermore, in a recent 
paper by Ansorg {\etal} \cite{AnsKM03a} a sequence of compressible 
rings is investigated with high precision. A bifurcation occurs 
from the Maclaurin sequence to the multi-ring sequence. 
What remains to be explored is the case of a full multi-ring 
sequence and its limit for the number of rings approaching
infinity. An interesting connection could appear with the
theory of disc fragmentation into individual rings \cite{GolL65}. 
The theory of Maclaurin spheroids and rings is 
built from simple arguments and is fundamental for 
understanding the problem of gravitating discs, 
but it still needs to be enriched with other physical ingredients.

Disc modelling poses a difficult task even in the Newtonian regime. 
In addition to the technical embarrassment of solving the Poisson
equation with sufficient accuracy,
difficulties are connected with strong coupling between 
magnetohydrodynamic terms and radiative transfer in the medium, 
and with our ignorance of boundary 
and initial conditions (they cannot be sufficiently 
constrained by observations but are decisive for the evolution of 
flows).\footnote{Note that even in the standard (non-self-gravitating) thin
disc model, the black-body temperature of the disc medium does not scale 
proportionally to the central mass. The enormous diversity 
in opacity, spatial and time scales, etc.\ does not allow for bare
scaling between corresponding physical quantities from one system to 
another. As a result, {\em{}Galactic {\bh} candidates 
have rather different discs from those around super-massive 
{\bh}s in galactic nuclei.} Distinctions and common properties of 
these two fiducial cases of accretion flows have been reviewed 
by Mirabel \cite{Mir03} and Czerny \cite{Cze03}.} Furthermore, 
huge values of the Reynolds number (typically $\gtrsim 10^{15}$) 
are associated with accretion flows. In consequence, simulations 
do not currently reach the numerical resolution required to
investigate simultaneously both the large-scale streams and 
the smallest swirls where kinetic energy is dissipated. This is another 
reason why (magneto-)hydrodynamical simulations of accretion
flows should be interpreted with caution \cite{Lon02}. Despite the 
tremendous complexity of these problems, models and 
simulations have reached a fair level of credibility \cite{HawBW99}.
Thermal equilibrium of the disc and its possible fragmentation
have been also explored \cite{Gam01}.

Given the technical obstacles inherent in the problem, disc modellers
often resort to a semi-phenomenological parameterisation of the turbulent 
transport by using a viscosity coefficient of the form
$\nu_{\rm{}t}={\alpha}c_{\rm s}H$, where $H$ is the disc half-thickness and
$\alpha\le1$ is a parameter. The $\alpha$-parameterisation,
originally advocated in the context of binary accretion discs by Shakura \&
Sunyaev \cite{ShaS73}, remains a recipe enabling study of the
characteristics of a mean accretion flow at the lowest level of
approximation. However, it does not explain the process of
triggering and maintaining the turbulence. Other viscosity laws have been
proposed, for instance the $\beta$-prescription which is based on
analogy with turbulence observed in laboratory sheared flows and
gives $\nu_{\rm{}t}\,\propto\,v_\phi R$
\cite{LynP74,RichZ99}. Also, there is the prescription of
marginally stable ($Q=\mbox{const}$) discs with
$\nu_{\rm{}t}\,\propto\,v_\phi{H^2}/R$, designed for weakly (locally)
self-gravitating discs \cite{BarHD98}. 
Very probably turbulent transport and its effects cannot be
satisfactorily mimicked through simple prescriptions.
Below we touch on several properties
which can be derived from the classical $\alpha$-prescription, as well as
the properties of $Q=\mbox{const}$ discs. 
Given the current knowledge, one cannot decide which
of these approximations is better.

\begin{figure*}
\begin{center}
\includegraphics[width=0.49\textwidth]{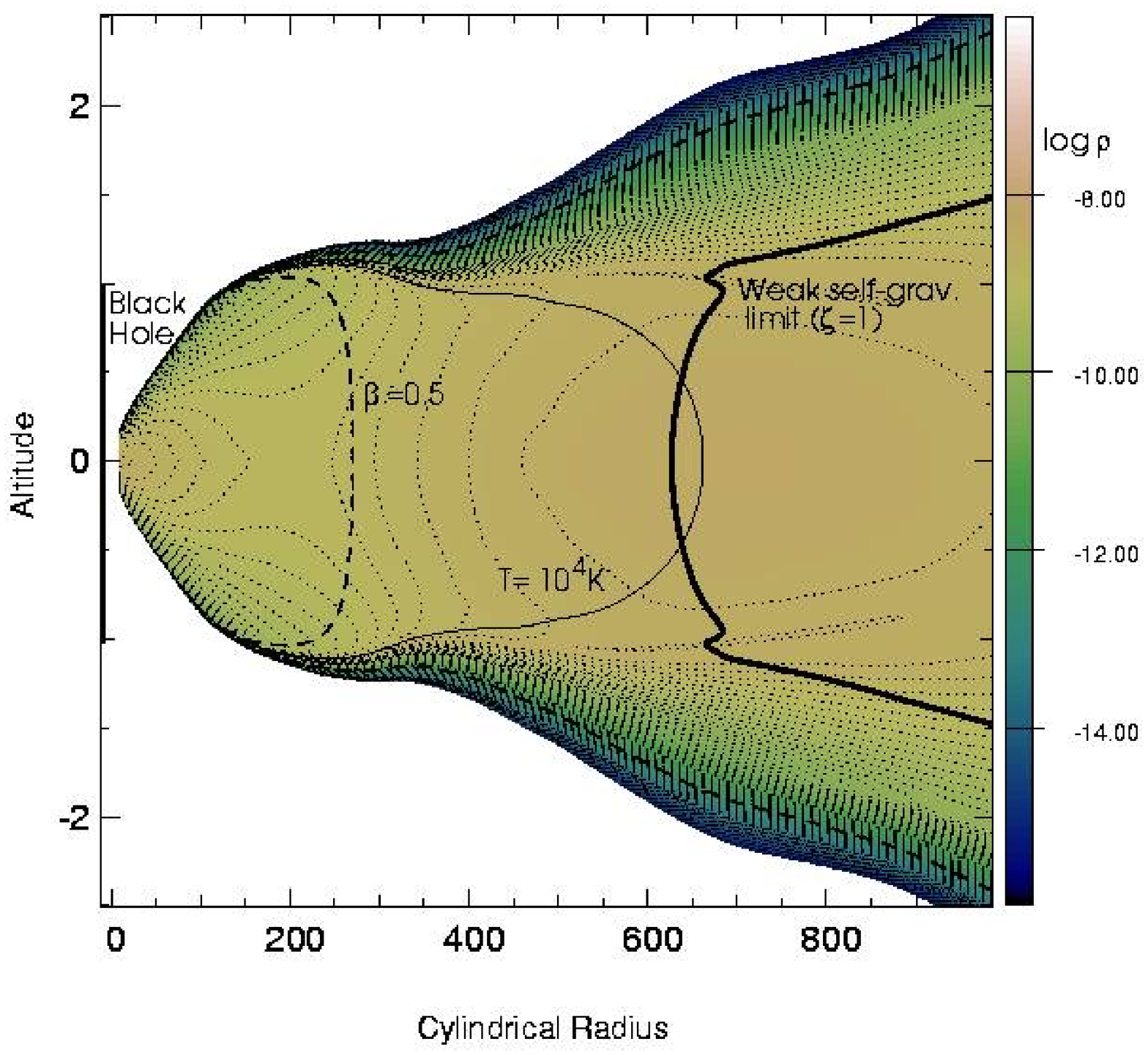}
\includegraphics[width=0.49\textwidth]{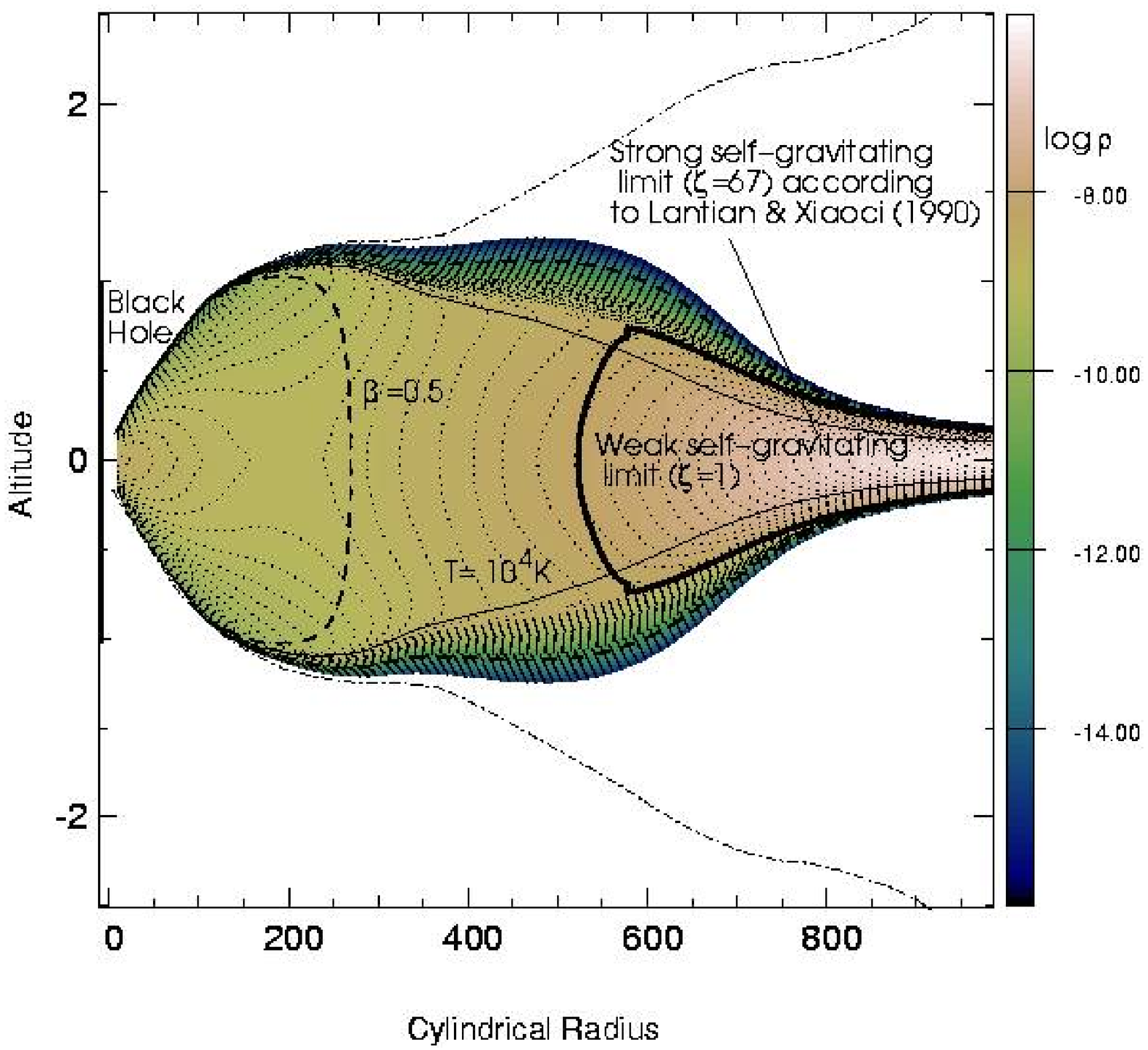}
\caption{The shape and density distribution of an
accretion disc feeding a $10^8M_\sun$ black hole. Left: the
\wsg\ limit (cf.\ section~\ref{sec:weak}) is indicated by 
$\zeta(R,z)=1$ (thick curve).
The limit is reached in the equatorial plane as close as $\sim 600\rs$.
At this point and beyond, the disc self-gravity is expected to play a
dominant role in the structure and dynamics of the flow
(the \ssg\ limit and the unstable limit are out of range adopted in this
plot). Right: The case of the \ssg\ limit is shown.
A standard-type steady-state model was employed \cite{Hur00,Pri81}
(the viscosity parameter is $\alpha=0.1$ and the accretion rate
$\dot{M}=0.1M_\sun\,$yr$^{-1}$). Lengths are plotted linearly and
normalised to the Schwarzchild radius; the density scale is in g/cm$^3$.}
\label{fig:map}
\end{center}
\end{figure*}     

\subsection{Poisson's equation in discs}
\label{sec:technical}
Association of the density field $\rho(\vec{r})$ with a prescribed
Newtonian potential is straightforward and always possible. 
Various classes of density--potential pairs can
be formed \cite{BinT87,EvaZ92}. Unfortunately, the reverse procedure,
i.e.\ the inversion of Poisson's equation, is difficult in
general. No analytically tractable expression is available for $\Psid$
that would correspond to a general three-dimensional system of finite 
size and mass. 
Exact solutions corresponding to non-spherical systems  
cannot be written generally in a closed form if the matter
distribution is two-dimensional, non-homogeneous or
of non-trivial shape or topology. Astrophysical discs exhibit 
all three complications. 

As an illustration we show first, in figure~\ref{fig:map} 
(left panel), the density distribution in the
standard disc model where self-gravity effects are {\it{}not\/}
included \cite{Hur00}. The complexity of shape and
density distribution is striking: discs appear to be complex
objects for which the merit of special analytical 
(and especially self-similar) solutions is very limited or 
doubtful. 

An alternative approach is to consider the integral definition
for the potential,
\begin{equation}
\Psid(\vec{r}) = - \int_{\rm disc}
 {\frac{\gconst\rho(\vec{r'})}{|\vec{r'}-\vec{r}|}}
 \;{\rd}^3{r'},
\label{eq:poisson-integ}
\end{equation}
and to use expansions of Green's function (for instance in
cylindrical harmonics or another kind of equivalent infinite series
representation). This approach is often employed in order to find 
self-similar solutions and to investigate the disc stability 
\cite{Hun63,LauR96}. However, as pointed out in \cite{Cle74}, 
convergence of the series is not guaranteed in all points of matter 
distribution. Also, the expansion leads to uncertainties because the 
inferred series must be truncated.

In order to be able to treat realistic systems, one has to seek
numerical solutions. However, it is difficult to achieve 
sufficient computational accuracy using standard techniques. Why
is this? The first reason is that no boundary conditions are 
available in the vicinity of
the disc; instead, they can be imposed only at infinity 
($\Psid\rightarrow0$ for $r\rightarrow\infty$). Secondly,
the physical boundary of the fluid is not known
{\it a priori\/} and it can evolve with time. One cannot rely on 
the standard situation that is met within laboratory physics of a fluid
confined within walls, in which case the boundary conditions are imposed 
by experimental settings. The third difficulty is caused by the relatively
large and variable aspect ratio: $H/R$ may be as large as unity and even 
greater at some parts of the disc, and as low as $\sim 10^{-5}$ at
other places (cf.\ figure~\ref{fig:map}). 
This fact hinders discussion of discs in active galactic
nuclei, where the geometrical thickness of the disc can vary wildly over
its diameter.

There are two practical ways of solving the problem
numerically \cite{MulS95}. The first one employs 
straightforward determination of the potential from
equation (\ref{eq:poisson-integ}) everywhere in the disc. 
This approach can be very accurate if one is able to
cope with integrable singularities inside the source
($\vec{r}\rightarrow\vec{r'}$) and it can
account for very distorted disc shapes 
and large density gradients. However, the computations 
involved are generally
very time consuming. One can envisage some cell-based
technique \cite{Sta83} (as done in particle
simulations \cite{HocE88}), but this is not expected to reach 
high precision (which is necessary for stability analysis).
Another possibility is to invert equation (\ref{eq:poisson})
using a finite difference method or a finite element method
\cite{Sto93}.
This approach requires that boundary conditions can be obtained
through equation (\ref{eq:poisson-integ}). 
Finally, it is worth noting that the inversion of Poisson's equation
for rotating stars can be performed via the mapping technique, 
e.g.\ transformation of space coordinates 
\cite{BonGM98,Cle74}. To our knowledge, 
the mapping technique has never been used for discs.

Since the efficient numerical computation of the potential is not trivial,
people still seek approximate
semi-analytical approaches that would complement
purely numerical solutions and make them faster. 
This is in particular the case for the infinite slab
(\isl) approximation, which was introduced by Paczy\'nski and
collaborators in the late 1970s in a series of papers
devoted to marginally stable discs
\cite{KozWP79,Pac78a,Pac78b}.\footnote{See also numerous follow-up papers
which helped to clarify substantial uncertainties in the original
formulation: \cite{BarHD98,BodC92,CanR92,Hur98,HurCLP94,Hur00,LiuXJ94,
PanG88,SakC81,ShoW82,ShlB87,ShlB89,ShlBF90,Sto93}. In particular,
the role of self-gravity and the disk fragmentation in fuelling the central
{\bh} has been discussed, but this topic remains beyond our discussion.}
In this approach the disc is examined in $(R,z)$-plane,
imposing axial symmetry and omitting the radial and azimuthal terms in
equation (\ref{eq:poisson}). The Poisson equation then reads
\begin{equation}
\frac{\partial^2 \Psid }{\partial z^2} = 4 \pi \gconst \rho.
\label{eq:psiisa}
\end{equation}
The main advantage of equation (\ref{eq:psiisa}) is the possibility of
determining the magnitude of the vertical gravity by direct integration
along the vertical direction. In this way even a full solution
$\Psid(R,z)$ could apparently be found. The vertical field is given by
\begin{equation}
g^{\rm d}_z(z)=-\frac{\partial\Psid}{\partial z}=-4\pi\gconst\Sigma(z),
\label{eq:gzisa}
\end{equation}
where
\begin{equation}
\Sigma(R,z) = \int_0^z{\rho(R,z')\,{\rd}z'}
\label{eq:sigmadef}
\end{equation}
is half of the local surface density. This result is
exact if the system is invariant under any translation
perpendicular to $z$-axis, and for this the disc must be
(i)~of infinite size, (ii)~strictly flat (its geometrical thickness
$2H$ is constant) and (iii)~homogeneous in the radial direction
(vertical gradients can be non-zero in the direction of $z$-axis). 
Obviously, such properties are not realistic, but the
validity of the \isl\ approximation can be extended to
radially inhomogeneous, finite size discs with non-constant
thickness, provided that the condition
\begin{equation}
\frac{1}{R}\frac{\partial }{\partial R} \left( R\,\frac{\partial \Psid
}{\partial R} \right)  \ll  \frac{\partial^2 \Psid }{\partial z^2}
\label{eq:isaextended}
\end{equation}
is satisfied. It is not generally possible to check
inequality (\ref{eq:isaextended}) for a given configuration because the
``true'' potential $\Psid$ is not known or is difficult to compute
with sufficient accuracy (for the technical reasons mentioned above).
Also, the above-described way of getting $\Psid(R,z)$ 
cannot be used to estimate 
the radial component of the field.

It is worth noting that equation (\ref{eq:gzisa}) holds
regardless of the relative masses of the disc and the central object,
making the \isl\ approximation often very useful (for example,
it can be used to compute massive non-Keplerian discs), but it
is also important to know where this powerful approximation
fails. Although the potential is a global function, we can determine 
this by considering its local properties by expressing
$\Psid{\propto}R^{a_1} |z|^{a_2}$.
Inequality (\ref{eq:isaextended}) is violated in regions where
\begin{equation}
\frac{|z|}{R} \gg \sqrt{\frac{a_2(a_2-1)}{a_1^2}}\,,
\label{eq:isa-psiab}
\end{equation}
or, in terms of density gradients, when
\begin{equation}
\left( \nabla_{\!R}\, \rho \right)^2 \gg \left( \nabla_{\!z} \rho +
z^{-1}\rho \right) \left( \nabla_{\!z} \rho + 2z^{-1}{\rho}\right).
\label{eq:isa-psirho}
\end{equation}

For a reasonable density stratification and attractive potential, 
we expect $a_1<0$ and $1<a_2<2$. This implies that the \isl\
approximation may break down when (i)~$a_1{\sim}a_2$, which corresponds
to discs and tori of moderate and large aspect ratios, and/or 
(ii) the radial and vertical density gradients are of the same order. 
This applies if density waves or shocks with steep
fronts are present, especially in boundary layers or close to the disc
edges. The flatter is the disc, the weaker are the edge effects.

\begin{figure*}
\begin{center}
\includegraphics[width=0.4\textwidth]{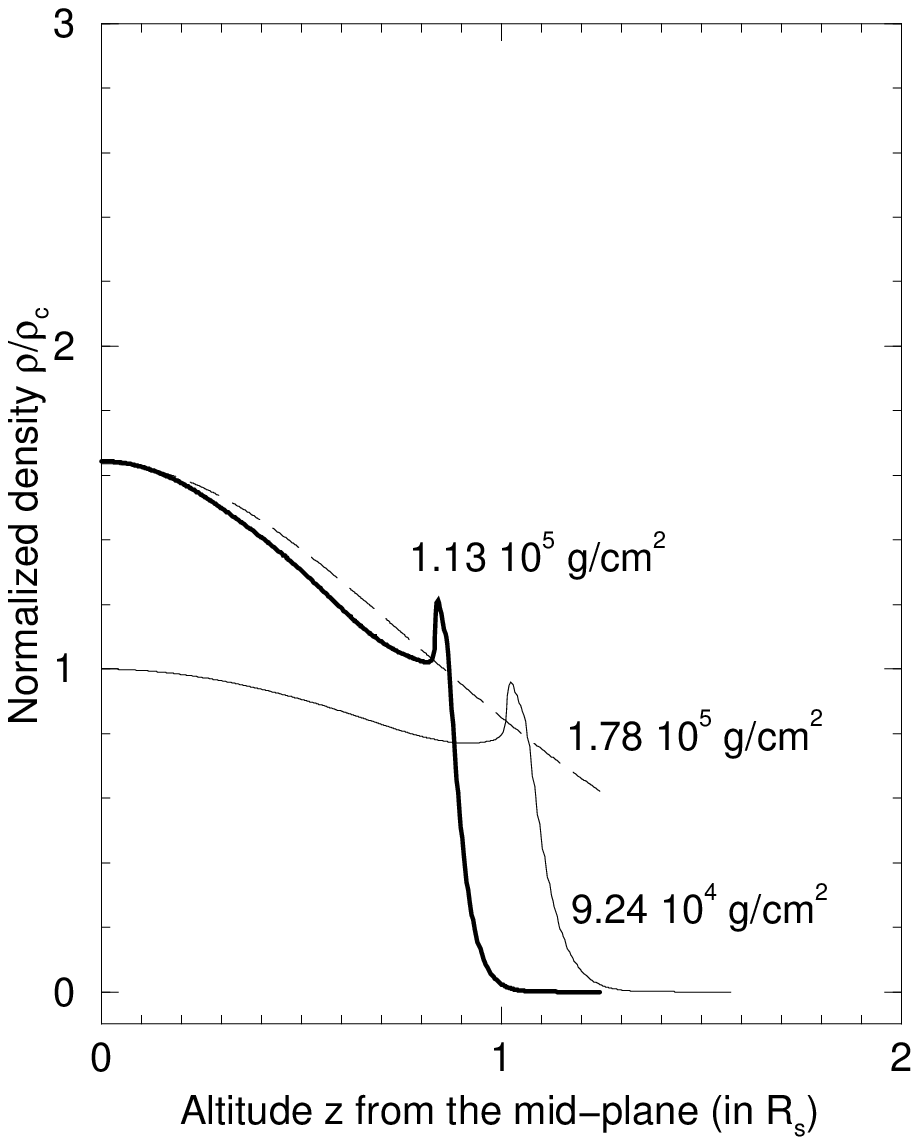}
\hspace*{5mm}
\includegraphics[width=0.4\textwidth]{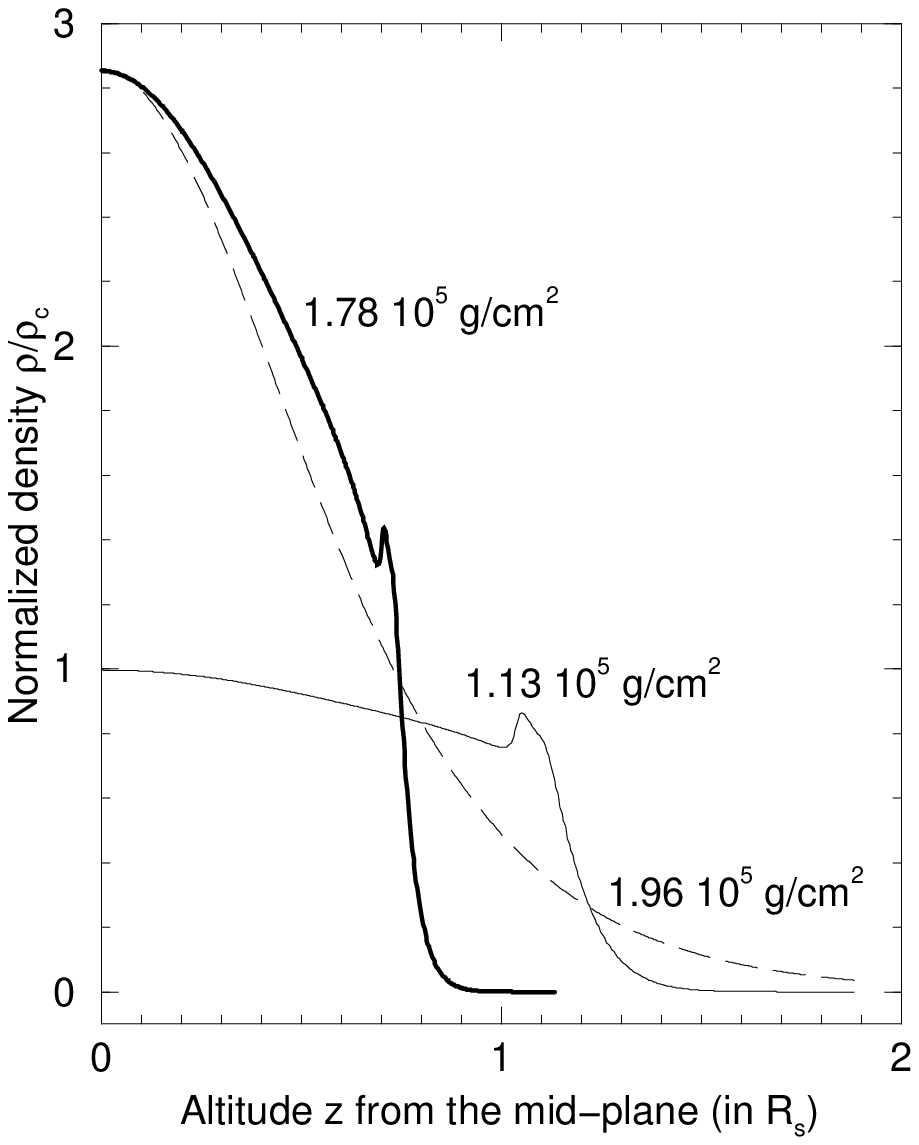}
\caption{The vertical distribution of density in a standard-type disc (same
conditions as for figure~\ref{fig:map}) at distance $r=500\rs$ 
(left panel) and  $r=600\rs$  (right panel). Three cases are shown,
all with $\mbh=10^8M_{\sun}$: (i)~In the first case (thin solid line),
hydrostatic equilibrium is maintained by the central
object while the disc self-gravity is deliberately neglected. The
mid-plane density is then equal to $\rho_{\rm c}$.
These calculations include vertical heat transfer by radiation and
convection. (ii)~In the second case (thick solid line), the disc
vertical self-gravity is introduced in the \isl\
approximation \cite{Hur00}. Despite the evident accumulation of
matter closer to the disc mid-plane, the area under the curves 
(which corresponds to half the surface density $\Sigma(H)$, as
labelled on each curve) is almost unchanged. Details of
stratification are sensitive to the mode of energy
dissipation. (iii)~Also shown is the solution for a
self-gravitating disc in the approximation of the Lane-Emden equation 
with isothermal structure (as given by equation (\ref{eq:solfk}); 
dashed line).}
\label{fig:rhoz}
\end{center}
\end{figure*}

An interesting property of vertical stratification can be derived 
directly from the \isl\ approximation \cite{FukS92,Pac78a}.
If the disc is able to reach hydrostatic equilibrium, then
the total pressure gradient compensates total gravity. 
Differentiating equation of hydrostatic balance relative to $z$ gives 
\begin{equation}
\frac{\rd }{\rd z}\left[\frac{1}{\rho}\frac{\rd P }{\rd z} +
\frac{{\gconst}Mz}{(R^2+z^2)^{3/2}}\right]+ 4\pi \gconst \rho = 0.
\label{eq:hydro}
\end{equation}
This equation becomes independent of the radius in 
the limit of a disc that stays in vertical equilibrium under its own 
gravity. An analytical solution can be obtained assuming 
polytropic relation $P = c_{\rm{}s}^2 \rho^{1+\frac{1}{n}}$
($c_{\rm{}s}$ is a speed, $n$ is polytropic index). The above-given equation 
is analogy of the famous hydrostatic equilibrium problem for gaseous 
spheres in equilibrium. The resulting ``plane'' Lane--Emden equation writes 
\cite{GolL65}
\begin{equation}
\frac{1}{\rho}\,\frac{\rd}{{\rd}z} \left(\rho^{\frac{1-n}{n}}\,
\frac{{\rd}\rho}{{\rd}z} \right) +
 \frac{4\pi{\gconst} n}{c_{\rm{}s}^2\left(1+n\right)}=0.
\label{eq:ple-polytrope}
\end{equation}
For vertically isothermal structure with $n\rightarrow\infty$, 
the equation (\ref{eq:ple-polytrope}) is the Frank--Kam\'enetski equation 
\cite{Schm67}
\begin{equation}
\frac{{\rd}^2 \ln \rho}{{\rd}z^2} +
 \frac{4\pi {\gconst} \rho n}{c_{\rm{}s}^2\left(1+n\right)}=0
\label{eq:fk}
\end{equation}
and the solution is \cite{GolL65,PanG88,SakC81}
\begin{equation}
\rho(z)=\rho_{\rm c}\,{\rm cosech}^2 \left(zc_{\rm s}^{-1}\,
\sqrt{2\pi {\gconst} \rho_{\rm c}} \right) ,
\label{eq:solfk}
\end{equation}
where $c_{\rm s}$ is the isothermal sound speed and 
$\rho_{\rm{}c}\equiv\rho(0)$ is the mid-plane density. From
equation (\ref{eq:sigmadef}), the surface density of the 
self-gravitating disc is given by 
$\Sigma = \sqrt{2 P_{\rm c}/\pi {\gconst}}$
\cite{PanG88},
and thus the disc mass depends solely on the radial distribution of central
pressure $P_{\rm c}=\rho_{\rm c} c^2_{\rm s}$. Discussion of solutions
for various polytropic indices can be found in
\cite{IbaS84,Pac78a}. The issue is that, as the polytropic index
$n$ runs from infinity (in the isothermal case) to smaller values, the
vertical density profile tends to be flatter. In
the limit $n \rightarrow 0$, this profile becomes exactly rectangular in
shape. Figure~\ref{fig:rhoz} shows the density profile in the direction
perpendicular to the disc plane in a typical disc. The isothermal case
is also shown for comparison. Models of the disc vertical stratification
are close to the models of 
stellar interiors and show that the vertical structure is not isothermal
and cannot be modelled by a polytrope (with $n=\mbox{const}$), at
least not if the disc is optically thick. Discs are heated by turbulent
motion and this results in a very complex vertical stratification of
physical quantities.

\subsection{What criterion for self-gravity in thin discs?}
\label{sec:weak}
In many respects it would be useful for the disc modelling
if the degree of self-gravity were indicated by an
unmistakable criterion. 
As a relevant quantity, one can intuitively
suggest the mass ratio $q$ $(=\md/\mbh)$, where
\begin{equation}
\md(R) = 4 \pi \int^R{\Sigma(R,H)R'\,{\rd}R'}
\end{equation}
is the cumulative mass of the disc and $\Sigma(H)$ is a half of the local
surface density according to equation (\ref{eq:sigmadef}). One might
postulate that self-gravity is unimportant for $q\lesssim10\%$, say.
Note that, in a binary system, the $q$-value (defined as 
the mass ratio of the two stars) is critical in the definition of 
the Roche lobes. However, such a criterion is unreliable
to explore an extended disc in its close 
neighbourhood, where it cannot be reduced to a point mass. 
We show below that the right criterion is based on the
quantity $\sim qR/H$ which contains information about
the disc mass (relative to the central black hole) as well as
about the distribution of the mass density in space
(in terms of the aspect ratio).

The ratio $\zeta=g_z^{\rm d}/g_z^\bullet$ can be evaluated
provided that
the vertical component of the disc self-gravity is known.
For $\zeta=1$, the disc and the central body contribute
comparably to the field along $z$-axis 
(this case is referred as the ``weak self-gravitating limit'', 
\wsg\ hereafter).
As soon as $\zeta$ exceeds unity, the disc is {\it vertically\/} 
self-gravitating. Within the \isl\ approximation, $g_z^{\rm d}$
is given by equation (\ref{eq:gzisa}), so one obtains
$\zeta=4\pi\Sigma(R,z)r^{3}/(\mbh{}z)$.
When the gas remains confined within altitudes 
$z/R\ll1$, we can use the mid-plane value
\begin{equation}
\zeta_{\rm c} \sim \frac{4 \pi \rho_{\rm c}R^3}{\mbh},
\label{eq:zeta}
\end{equation}
where $\rho_{\rm c}$ is the mass density in $z=0$,
as defined in the absence of self-gravity. The role of parameter
$\zeta(R,z)$ was discussed in \cite{Hur98};
Taylor expansion around $z=0$ shows that the mid-plane value of
$\zeta$ can be identified with $A$-parameter of Paczy\'nski 
\cite{Pac78a}. One finds
that the upper limit for the disc height is typically $H\lesssim0.3R$; 
such discs stand on a borderline beyond which the approximation of
geometrically thin discs becomes very crude and inadequate. We notice
that vertical self-gravity occurs preferentially (i)~when the central
{\bh} has a small mass, (ii)~if the local mass density in the
disc is large, and (iii)~at rather large distances from the centre. If
the surface density obeys a power-law in radius, for instance in 
the form $\Sigma=\Sigma_0 R^s$, then
\begin{equation}
\zeta_{\rm c} = q(R)\, (s+2) \,
\left( \frac{H}{R}\right)^{-1} \qquad (s \ne -2).
\label{eq:zetac}
\end{equation}
Relation (\ref{eq:zetac}) shows that even low-mass discs can 
be self-gravitating
if they are flat enough. We can put numbers in this relation: given
that standard discs in active galaxies have a power 
index $s \sim -0.6$ and small, 
almost constant aspect ratio, $H/R \sim 0.01$, we find $\zeta_{\rm c} =
140 \times q(R)$  \cite{Hur98}. (The adopted value
of $s$ corresponds to the outer, gas-pressure dominated region;
another value would not change qualitative conclusions at this point.)
In other words, 
a disc becomes vertically self-gravitating when its relative mass 
exceeds a small threshold value. This value is model dependent; 
typically it is order of a few $\times10^{-2}$. Note that the limit is
much less than the crude estimate of $10\%$ suggested before.

By introducing ratio 
$\eta\equiv\nabla_{\!R}\Psid/\nabla_{\!R}\Psih$
one can follow the same kind of analysis for the radial
component of gravity.
A disc is {\it radially\/} self-gravitating if $|\eta| > 1$, hence
this regime can be referred to as a strongly self-gravitating (\ssg) disc.
Conclusions are less straightforward in this case because we lack
sufficiently reliable expression for
$g_R\equiv-\nabla\Psid$. In other words, there is no obvious
analogy to the \isl\ approximation. Some
authors \cite{MinU96,Tsu99} have proposed that the total
potential can be approximated by the monopole term, 
$\Psid \sim -{\gconst}\md(R)/R$, which is not fully satisfactory but 
gives at least the right order
of magnitude in the \ssg\ limit ($\eta\sim1$).
Considering, as above, a power law for the radial surface density
profile, $\eta$ is given by
\begin{equation}
\eta_{\rm c} = q(R) - \zeta_{\rm c} \frac{H}{R} = (1+s) q(R)
\label{eq:etac}
\end{equation}
in the equatorial plane. Using equation (\ref{eq:zetac}), we 
find
\begin{equation}
\eta_{\rm c} = \zeta_{\rm c} \frac{1+s}{2+s}\frac{H}{R}
\end{equation}
(the case of $s=-2$ can be derived in similar manner). 
The \ssg\ regime corresponds to
\begin{equation}
\zeta_{\rm c} = \frac{2+s}{1+s} \left( \frac{H}{R} \right)^{-1} \gg 1.
\end{equation}
Using the above values for $s$ and $H/R$, $\eta_{\rm c}=1$
corresponds to $\zeta_{\rm c} \sim 150 $, well above the \wsg\
limit. As $\zeta_{\rm c}$ is expected to be an increasing
function of radius, the \ssg\ limit involves more mass and is
reached at larger radii than the weak limit. It therefore lies inside the
vertically self-gravitating discs. This value is in good agreement
with the results established by Lantian \& Xiaoci \cite{LanX90} who
have attempted to performed a direct, analytical integration of
equation (\ref{eq:poisson-integ}). Although their investigation still
contains some rough approximations (for instance, singularities in
the Poisson kernel were artificially removed), these authors
derived a formula for radial self-gravity which 
accounts for finite size effects. For $s=-1$, they found radial
acceleration to be
\begin{equation}
g_R^{\rm d} = 2 \gconst \Sigma\left[ \ln \frac{(2R+H)(R+R_{\rm
in})(R_{\rm out} -R)}{(2R-H)(R-R_{\rm in})(R_{\rm out}+R)} -
\frac{H^2}{4R^2}\right], 
\label{eq:lantian}
\end{equation}
which sets the \ssg\ limit at $\zeta_{\rm c} \sim 67$ (assuming the
Eddington accretion rate; see figure~\ref{fig:map} right). Note that
this formula still remains singular at the disc edge \cite{Mes63}. 
To circumvent the difficulty, a softening parameter is often
introduced \cite{LauKA98}, however,
correct behaviour can be obtained just by careful treatment of edges.

Results discussed so far have not yet addressed the essential
questions concerning the gravitational stability of 
self-gravitating, geometrically thin discs. It is known for long,
and confirmed by observations, that galaxies (treated as discs of 
stars, gas and dust) are regulated by self-gravity \cite{BinT87,Too77}. 
Similar processes can form spirals and bar 
patterns in purely gaseous discs. 
It was shown in a remarkable paper by Goldreich \& Lynden-Bell 
\cite{GolL65} that instability sets in when
\begin{equation}
\frac{\pi \gconst \rho_{\rm c}}{4 \Omega^2}\gtrsim\xi_{n},
\label{ineq:glb}
\end{equation}
where the lower limit, $\xi_n$, depends on polytropic index $n$ (the
value of $\xi_n=1.1$ corresponds to a vertically isothermal disc, 
while $\xi_n=1.75$ for $n\rightarrow0$). 
Note that this is an analogy of Toomre's original criterion,
but with a slightly different threshold 
\cite{Gam01,Goo02,PolPS97,Too64}. It
is important to recall that Goldreich \& Lynden-Bell's criterion has
been established from an analysis of growing unstable modes through a
linear perturbation theory, within the framework of \isl\ approximation
and assuming a non-viscous, uniformly rotating, flat disc. It is
therefore expected to be somewhat altered when relaxing some of the
above assumptions. It is easy to recognise that equation
(\ref{ineq:glb}) also gives $\zeta_{\rm c}\gtrsim18\mbox{--}28$. 

We thus conclude that instabilities are not present in a disc if its size
remains below the weak limit. The region separating the weak, the
unstable and the strong limit can be narrow or wide, depending on the
surface density profile; or, more precisely, on the function 
$\zeta_{\rm{}c}(R)$. Again, a very nice example is provided by
Saturn's rings which are vertically self-gravitating at their outer
edge, but gravitationally stable, implying that the weak limit and
the unstable limit do not coincide.

\subsection{The self-gravitating regime in accretion discs in active galaxies}
According to equation (\ref{eq:zeta}), an increase of the $\zeta$-parameter
is expected as the radius increases. This means that there is a
boundary zone that separates the inner (non-self-gravitating) 
solution from the outer (self-gravitating) one. The location of
this zone can be found without accounting for self-gravity, simply 
by computing $\zeta$ from a disc model \cite{ColD90,Goo02}. This approach
may underestimate the magnitude of self-gravity and overestimate the
radius of such boundary, but it provides a useful order of magnitude
estimate. A more reliable result must include vertical
self-gravity into consideration. This can be achieved within
the \isl\ approximation.
Furthermore, in the context of $\alpha$-prescription,  
the \wsg\ limit is reached at $R_{\rm{}sg}{\equiv}R(\zeta=1)$.
For accretion rates lower than a certain critical value, 
$\dot{M}_{\rm{}crit}$, one finds \cite{Hur98}
\begin{equation}
R_{\rm sg}\sim 480 \; \bar{\alpha}^{14/27}
\dot{M}^{-8/27} M_8^{-2/3} \bar{\kappa}^{2/9} \mu^{-8/9},
\label{eq:rsg1}
\end{equation}
while
\begin{equation}
R_{\rm sg}\sim 230 \; \bar{\alpha}^{2/9} \dot{m}^{4/9}
M_8^{-2/3}  \bar{\kappa}^{2/3}
\label{eq:rsg2}
\end{equation}
for accretion rates larger than $\dot{M}_{\rm crit}$; here,
$R_{\rm sg}$ is expressed in Schwarzschild radii,
and the critical accretion rate is expressed in
solar masses per year,
$\dot{M}_{\rm crit}\sim0.27\,\alpha_{0.1}^{2/5}\,\bar{\kappa}^{-3/5}
\mu^{-6/5}\;M_\odot$/yr ($\bar{\kappa}$ is opacity). 
Noticeable deviations are observed when considering temperature-
and density-dependent opacities and equation of state (not yet
specified self-consistently in equations (\ref{eq:rsg1})--(\ref{eq:rsg2})). 
The existence of two different relations arises from the 
fact that {\bh} discs generally possess an inner region
(dominated by radiation pressure) and an outer (gas pressure dominated) 
region where the physical quantities (namely, the density) exhibit 
different variations with radius and the accretion rate. The case 
(\ref{eq:rsg2}) corresponds approximately to accretion luminosity 
exceeding the Eddington limit.

\begin{figure*}
\begin{center}
\includegraphics[width=\textwidth]{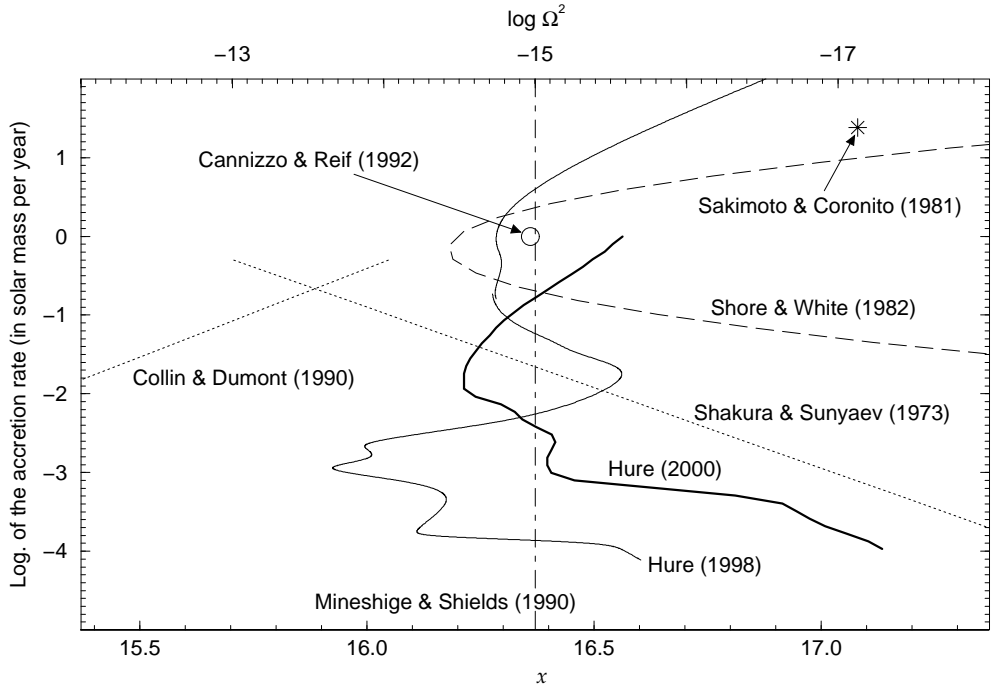}
\caption{The location of \wsg\ limit as a function of
the accretion rate in an $\alpha$-disc ($\alpha=0.1$). 
Top axis gives the logarithm of Keplerian 
orbital frequency ($\Omega$ in Hz), while the
bottom axis gives the corresponding value of 
$x\equiv{\log}R-\frac{1}{3}{\log}M_8$ (radius is in cm, the mass $M_8$ 
is in units of $10^8$ solar masses). The accretion rate is on the vertical 
axis. Individual curves correspond to different models as 
indicated. The \wsg\ limit was computed without including vertical 
self-gravity in all models except Hur\'e (1998, reference \cite{Hur98};
and 2000, reference \cite{Hur00}). The case \cite{Hur98} is one dimensional 
while \cite{Hur00} is two-dimensional. Distortions in the curves are 
given mainly by opacity which shows drastic variations with temperature.
A reference power-law line for the standard Shakura \& Syunaev
(1973, \cite{ShaS73}) non-self-gravitating disc is also plotted.}
\label{fig:weaklimit}
\end{center}
\end{figure*}

Figure \ref{fig:weaklimit} gives a theoretical relationship between 
the location of \wsg\ limit and the accretion rate. It was computed
by different authors using the $\alpha$-prescription 
and the same input parameters (relevant for accretion flows
onto super-massive {\bh}s in galactic nuclei). 
It is noticeable that no firm agreement has been achieved about the
position of this limit, which largely fluctuates from one author to
another simply because of different ingredients in models.
Also, there are no direct observations available to confirm if
this \wsg-transition operates in nature and if it occurs where predicted.
However, in active galaxies there is a {\em{}striking similarity} 
between the location and
size of the so-called broad-line region, and the value of $R_{\rm{}sg}$.
This correspondence could indicate that 
the clouds are formed at the distance where
self-gravity becomes important (cf.\ \cite{ColH01}
for a systematic discussion of the correlation). 
According to bi-dimensional calculations, 
a lower (reliable) limit appears to be $R_{\rm sg} \sim 530\rs$
for a $10^8$ M$_\odot$ central black hole and $\alpha =0.1$. This is
only 5 milliparsecs from the centre, much less than previously estimated
(e.g.\ \cite{ShlB87}).  

\subsection{One dimensional models: self-similar and asymptotic solutions}
\label{oned}
Of simple models that can be constructed analytically on the basis of
$\alpha$-parameterisation, we pay a special attention to those where
self-gravity is taken into account in the vertical direction,
while Keplerian rotation is still imposed. This represents even
a weaker category of {\wsg} discs, because their rotation curve is 
assumed not to be altered by the disc mass. Within this approach, 
the above-discussed infinite slab approximation has
been found very convenient, although some criticism can be raised at this 
point, because these disc models typically tend to be rather massive.
Hence they become globally self-gravitating and no more Keplerian
(this problem occurs especially if the surface density falls 
faster than the square of radius).
Paczy\'nski's original paper \cite{Pac78a}
was devoted to the investigation of discs with constant 
ratio of the vertical self-gravity to the vertical
component of the central field, i.e. constant value of
$\zeta$-parameter (i.e.\ $\zeta = g_z^{\rm d}/g_z^\bullet$, which
corresponds to $A$ in Paczy\'nski's notation).
The assumption of vertical self-gravity is justified by the fact that strongly
self-gravitating discs are prone to gravitational instabilities
\cite{GolL65,Too64} which should make them evolve in a couple of
dynamical time scales ($t_{\rm dyn}\sim\Omega^{-1}$). The endpoint
of this evolution is the state of constant Toomre's $Q$-parameter,
which is typical for differentially rotating discs of stars
($Q$ can be related to $\zeta$ via equation (\ref{eq:zeta})).

Soon it was realized that holding $\zeta=\mbox{const}$ might produce
some substantial bias \cite{ShoW82}; a series of papers followed to
elucidate the local response of the disc to vertical self-gravity 
and to study $\zeta(R)$ profile. Keplerian discs 
were investigated in which $\zeta$ varies as a function of radius
in a self-consistent fashion (another motivation for this research 
was the chance of discovering new branches of solutions). 
Most explorations were carried
out for steady discs, using the so-called ``vertically averaged
approximation'', where one seeks only for radial profiles of density,
temperature and other physical quantities (without considering 
vertical stratification in detail). With additional assumptions 
about opacity, equation of state, pressure regime,
etc., one is naturally lead to self-similar solutions 
\cite{MedN00,MinU96,Tsu99}.

As already mentioned above, vertical self-gravity makes the accreted 
gas more concentrated around the mid-plane. As a result, the disc tends 
to collapse in the vertical direction beyond some critical radius where 
$\zeta\gtrsim1$ (cf.\ figures \ref{fig:map} and \ref{fig:rhoz}).
This property is shared by all computations of geometrically thin discs,
independently of the actual viscosity
prescription (in other words, if one assumes $\alpha$-parameterisation
for viscosity, then the disc height is almost independent
of $\alpha$). The opposite is true for
geometrically thick discs which tend to be thicker when self-gravity is
taken into account \cite{BodC92}. If prescription for viscosity
is specified, the variation with radius can be found in explicit form.
Examples are given in table~\ref{tab:r-variations};
given the viscosity law and the heating mode of
the disc that is on verge of self-gravitating regime, 
one finds the $R$-variation of the disc thickness, temperature, surface 
density, etc. Note that $\zeta(R)$ increases with $R$ in
all models. This leads inevitably to the onset of 
self-gravity, subsequent reduction of the disc thickness and 
further consequences on many aspects, such as 
self-irradiation, dynamical stability and the emitted spectrum 
of the disc. It may also affect the disc warping \cite{Pri96}
in regions that are shadowed from the central source and 
cannot reprocess its radiation; although weak at those distances,
the resulting lack of heating is in favour of gravitational instability.

\begin{table}
\begin{center}
{\footnotesize
\begin{tabular}{lccccc}
Model & $T_{\rm c}$ & $\Sigma$ & $H$ & $\zeta_{\rm c}$ \\
 \hline 
$\alpha$-disc (no self-gravity) \rule{0mm}{2.8ex} 
 & ${-9/10}$ & ${-6/10}$  & ${21/20}$  & $27/20$ \\
Self-gravitating $\alpha$-disc  
 & 0       & ${3}$      & ${-3/8}$  & $9$ \\
$\beta$-disc (no self-gravity)
 & ${-7/8}$  &  ${-1/2}$   & ${17/16}$ & $23/16$ \\
Self-gravitating $\beta$-disc  
 & ${-7/8}$  & ${-1/2}$  & ${-3/8}$  & $23/8$ \\
$Q$-constant disc              
 & ${-9/7}$  &  ${-15/7}$ &   ${6/7}$  & 0 \\ 
Irradiated $Q$-constant disc \rule[-1.2ex]{0mm}{2.6ex}   
 & ${-5/8}$  & ${-29/16}$ &   ${19/16}$ & 0 \\ \hline
\end{tabular}
}
\end{center}
\caption{In the \wsg\ regime, parameters of the disc are found 
proportional $R^s$ in the radial direction. 
In this table, the values of power-law index, $s$, are given
for different physical quantities: mid-plane temperature $T_{\rm c}(R)$,
surface density $\Sigma(R)$, geometrical thickness $H(R)$, and 
self-gravity parameter $\zeta_{\rm c}(R)$. The optically thick limit was assumed and
the $\alpha$-, $\beta$- and $Q$-constant prescriptions were adopted for 
turbulent viscosity. Self-similarity is maintained if gas pressure dominates 
in the medium with opacity coefficient and chemical composition being constant
\cite{BarHD98,HurCLP94}. Viscosity was incorporated e.g.\ by \cite{MinU96} 
in a self-similar time-dependent model. In spite of numerous
papers, it would be still premature to expect a general consensus 
regarding the viscosity model. For further details, see 
\cite{BodC92,CanR92,FukS92,HasEM95,KozWP79,LiuXJ94,Sto93,SakC81,ShoW82}.}
\label{tab:r-variations}
\end{table}

Properties derived through the $\alpha$-prescription deserve a
comment. The $R$-dependence of the $\zeta$ comes out 
rather steep and much more pronounced than what is obtained for
a bare standard disc (neglecting self-gravity), in which case 
$\zeta_{\rm{}c}(R)\,\propto\,R^{27/20}$. This means that the transition 
from non-self-gravitating regime to the fully self-gravitating 
one should occur in a very narrow region; the weak, unstable 
and strong limits are expected to almost coincide with each other. 
As a consequence of the steep rise of self-gravity, the disc 
can suffer from a runaway-type increase of density; one is even led to
densities typical for stellar interiors and conditions suitable 
for deuterium and hydrogen burning \cite{PanG88}. Temperature 
reaches a plateau in the outer region of the disc plane with value
\cite{DusSB00,Goo02,Hur00,ShlB87}
\begin{equation}
T = \mu\,m_{\mathrm H}\,k^{-1}\left(\textstyle{\frac{2}{3}}\;
 \alpha^{-1}\gconst\dot{M}\right)^{2/3}
 \simeq 2.4\times10^4\,\mu\,\alpha_{0.1}^{-2/3}\,
 \dot{m}^{2/3}\quad{\mathrm [K]}\,,
\label{eq_tsurmu}
\end{equation}
where $\mu m_{\rm H}$ is the mean mass per particle, $\dot{m}$ is the
accretion rate normalized to $0.1M_{\sun}/\mbox{yr}$ and
$\alpha_{0.1}=\alpha/0.1$. Some authors have argued that
such uniform mid-plane temperatures are unphysical and they
have suggested the use of a different prescription for viscosity 
\cite{BarHD98,DusSB00}. Interestingly, this may lead back to
the original idea of Paczy\'nski that viscosity is connected with
the $Q$-parameter.

\begin{figure*}
\begin{center}
\includegraphics[width=0.43\textwidth]{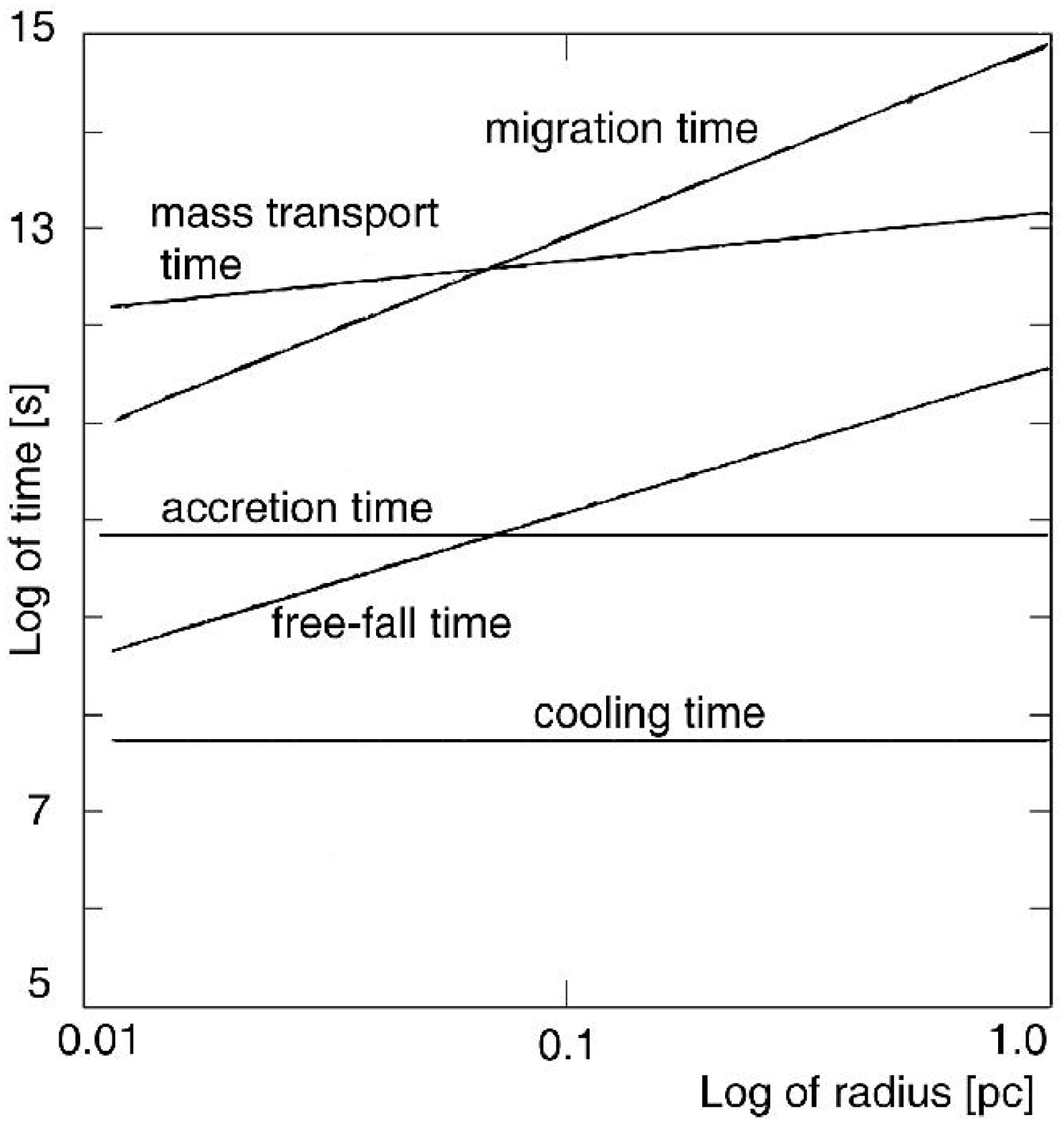}
\hspace*{1mm}
\includegraphics[width=0.47\textwidth]{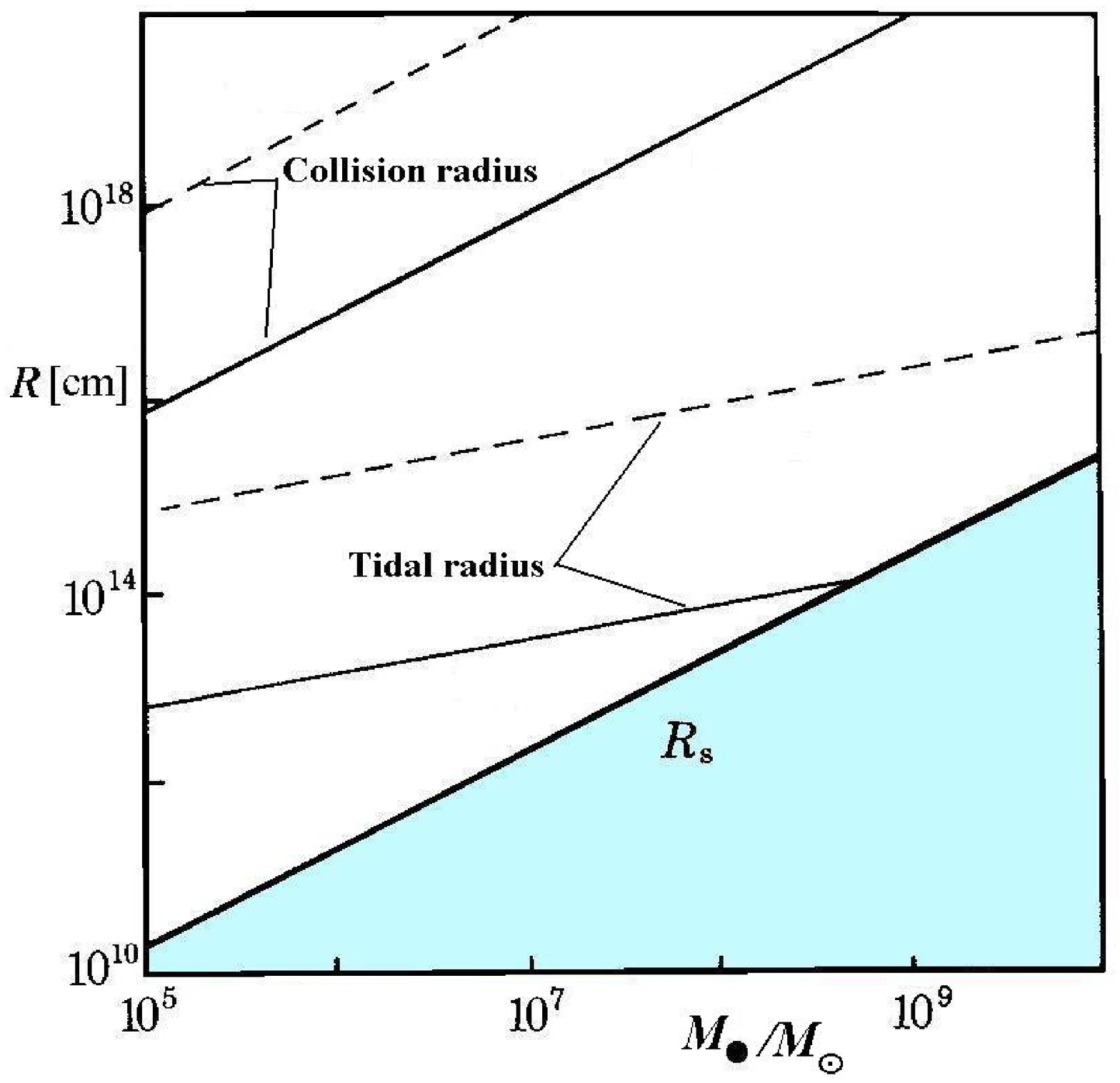}
\caption{Left: time-scales in the gravitationally
unstable region of an accretion disc \cite{ColZ99}. 
One can define the mass transport time in the disc, cooling time 
(relevant for the formation of fragments), accretion time 
(for a star growing to a pre-determined mass, $\mst\sim10M_{\sun}$), 
migration time (for a star sinking towards the centre), and free-fall 
time. Right: length-scales for a black hole embedded in a star cluster: 
the tidal disruption radius within which stars are disrupted, the 
collision radius within which colliding stars are disrupted, and
Schwarzschild radius ($R_{\rm{}s}$) of the central black hole.
Solar-type stars (solid lines) and solar-mass red giants (dashed lines)
have been assumed. Figure was adapted from \cite{Fab79,FraR76}.}
\label{fig:scales}
\end{center}
\end{figure*}

We conclude this section by showing several typical scales of the problem
in figure~\ref{fig:scales}. These graphs are derived from 
basic arguments and they can
help to identify physical mechanisms which are important from
the observational point of view. Time-scales relevant for the gravitationally 
unstable regions of accretion discs are shown as a function of radius 
(in the left frame). The possibility of star formation was taken 
into account together with continuous migration and evolution in the medium
(here, $\mbh=10^6M_{\sun}$ was assumed for the central black hole).
If the stars form a dense (quasi-spherical) nuclear cluster in 
which the central {\bh} resides, 
one can define the radius where stars are 
destroyed tidally or by their mutual collisions.
The resulting value depends on $\mbh$ as well as details 
of the stellar structure (as shown in the right frame).

\section{Waiting for observations and constraints}
\label{sec:consequences}
Significant effort has been devoted to searching for black holes 
during the last few 
decades. Traditionally, most approaches of present-day 
astronomy rely on the electromagnetic radiation coming from cosmic sources,
but there are other possibilities which will be surely exploited in
near future (neutrinos, high-energy cosmic particles, and gravitational waves). 
The idea of detecting black holes via (electromagnetic) radiation of 
nearby luminous matter can be traced back to the celebrated lecture, 
which rev.\ John Michell presented in front of the Royal Society as early as in 
1783.\footnote{A paper was based on that lecture and published under the title  
{\it On the means of discovering the distance, magnitude, \&c.\ of the fixed
stars, in consequence of the diminution of the velocity of their light\ldots}
in  Philosophical Transactions of the Royal Society of London, 1784 {\bf 74} 35.}
Understanding local physics of the medium is crucial if one aims to interpret
complex and variable spectra that are observed. X rays are particularly 
pertinent because they are produced near {\bh}, as discussed by Fabian and others 
\cite{Fab79}. Indeed, given the intrinsic 
emissivity of the medium, there are only a few factors determining 
the source spectrum on the Earth. The main factor is the black hole mass 
$\mbh$; the angular momentum of the black hole and the disc
self-gravity play secondary roles. The geometrical optics approximation 
is thus adequate and one can hope to infer actual values of {\bh} parameters
from the motion of test matter around it. However, one has to be well aware
of various environmental influences that hinder real observations.

\begin{figure*}[tb]       
\begin{center} 
\vspace*{1mm}    
\includegraphics[width=\textwidth]{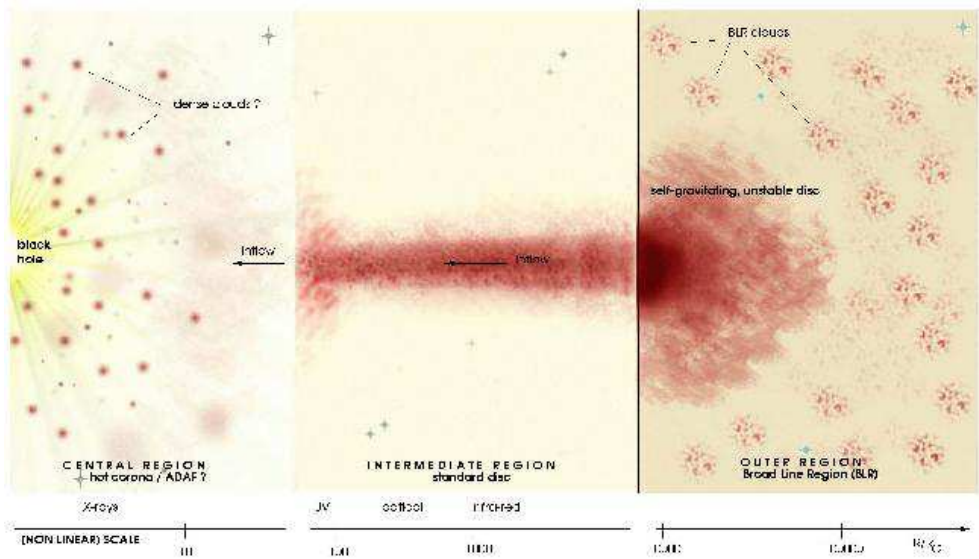}
\end{center}    
 \caption{A schematic drawing of the central part of a galactic nucleus. 
 Different regions of the accretion flow are shown: the innermost, 
 radiation-pressure dominated region with a central {\bh}, the intermediate 
 region with a standard disc, and the outer, gravitationally unstable part.
 Length-scales are proportional to the central mass. The indicated 
 values are valid for $\mbh=10^8M_{\odot}$ and slightly model-dependent
 \cite{ColH01}.}
 \label{fig:collin}
\end{figure*}

Under conditions that are thought to be typical in active galactic nuclei,
the inner regions of accretion flows contain diluted plasma with a 
corona threaded by magnetic fields \cite{Lei99,Pou01}. 
High-energy photons are first generated in the corona. 
These primary photons irradiate substantially colder gas of the 
accretion disc, where they are reprocessed further (figure~\ref{fig:collin}). 
A fraction of photons is reflected. The temperature of the
reflecting medium can be $\sim10^4\,$K in galactic nuclei and about
2--3 orders of magnitude higher in the case of Galactic {\bh}
candidates \cite{CasGP01,Don02}. The quality of photon reprocessing 
depends critically on the ionization state and other local
properties of the medium. In this way the resulting spectrum is formed
with spectral features (lines) superposed on top of the background
continuum. X-ray spectra contain precious information about rapid
flows from which one can infer properties of the central source
of gravity and the state of the nearby gas.

Manifestation of strong gravity has been searched in radiation of 
accretion discs with the main aim of finding the evidence for the
central black hole \cite{FabIRY00,Mar00,Rey96,ReyN03}.
Majority of these papers deal with non-self-gravitating discs
and assume that integrated light of the whole source is only very weakly 
affected by the disc gravitation (however, see \cite{KarLV95,SirG03,UsuNE98}). 
In the case of our Galactic centre, the presence of standard-type (thin)
accretion disc is rather improbable (e.g.\ Narayan \cite{Nar02}).
Direct effect of the disc gravity on spectral line profiles is below current 
observational resolution for any present-day spectra of active galactic
nuclei, so it can be 
safely neglected in the first approximation (we briefly discuss this effect 
in the following paragraph). It is noteworthy, however, that self-gravity
of the disc has indirect impact on spectral
features. First, as discussed in section~\ref{sec:astrophysical}, 
the disc geometrical shape is affected by self-gravity
with consequences for obscuration of some parts of the disc
and the resulting variability, especially at large values of inclination
(this is also the case when {\gr} effects of the central 
{\bh} are maximal, so the two influences are mixed). Self-gravity
also controls the amount of disc self-irradiation, which 
produces indirect signs in reflected light. Finally,
self-gravity induces fragmentation in the disc medium 
(depending in particular on the disc cooling properties \cite{Gam01}) with
fundamental consequences on the disc structure. Fragments can even
lead to star formation or contribute to star-bursts \cite{ColZ99,Elm02}.
The clumpy material may also result in much more pronounced variability 
observed from the source \cite{AbrC00}. 

Let us now discuss direct effects of the combined {\bh}--disc gravity 
on observed light rays. In the early works of Cunningham \& Bardeen 
\cite{Cun75,CunB72},
a standard non-self-gravitating disc was assumed. These authors 
initiated a series of investigations of model spectra with the aim 
of revealing the imprints of \gr. Since then, many people
have advanced various modifications of the original semi-analytical 
approach \cite{BaoWH98,BroChM97,FanCFC97,MarKM00,Vie93}. 
Very prominent role has been played by studies of
spectral features (lines) of iron in the X-ray band; 
in particular, the so called disc-line scheme 
has been widely applied, as it captures the main spectral characteristics
attributed to strong gravity \cite{FabRSW89,YouFRT01}. The line 
is produced by reprocessing the primary continuum radiation 
near above the inner edge of the disc,
within several tens of $\rs$ (see \cite{NayK01,RozDCC02} 
for recent exposition of emission mechanisms). 
The line intrinsic width is small, typically much less than the final 
observed profile that results from integration over the whole emitting 
surface. Standard picture of the disc thermal
spectrum \cite{NovT73,PagT74} was reconsidered to account also
for the disc gravity \cite{LaoN89}. 
For detailed examination it would be useful to produce templates of 
expected line profiles or even spectral models that are consistent with
the formation of non-axially symmetric structures in accretion discs 
as well as the presence of a black hole in the centre 
\cite{HarB02,KarMS01,UsuNE98}. There is an exciting
possibility of detecting gravitational effects 
on polarized light of a source orbiting a black hole, but this 
idea still awaits technology usable for its practical implementation 
\cite{ConSP80,Mat93}. Effects of the black hole 
itself on the light propagation are blurred in real sources
in which the signal comes from different, insufficiently resolved regions. 

So far, only a few cases have been discovered of an
accreting black hole with a line that attains the
characteristic shape, width and variability and could be
understood as the interplay of Doppler and gravitational shifts. 
The best example of such spectral feature originating near a {\bh} has
been identified in the nucleus of Seyfert galaxy 
MCG--6-30-15\footnote{Cf.\ Morphological Catalogue of Galaxies.} where the
iron line was caught at the moment of high redshift
\cite{Iwaetal96,Fabetal02}. Large gravitational
redshift can be connected with rapid {\bh} rotation, because
the radius of the disc inner edge (where the line presumably originates)
decreases with {\bh} angular momentum increasing 
\cite{Wiletal01}. Magnetic fields can change this straightforward
conclusion. Also, the line is strongly variable, so its final 
(continuum subtracted) shape cannot be resolved
reliably with current knowledge; but there is a good prospect of
overpassing these deficiencies with more detailed models and
using time-resolved spectroscopy that will
be able to track rapid features \cite{MerF03}. 
Reconstruction of accretion disc images will then be doable using 
reverberation techniques \cite{Pet01,Ste90} or Doppler mapping 
\cite{MarH88}. This will also require good understanding of
non-standard effects which are usually neglected, e.g.\
the effect of disc eccentricity \cite{BaoHO96,Sta01}.

As far as {\bh} candidates in the Milky Way are concerned, the relativistic 
iron line was detected in a well-known micro-quasar GRS 1915+105 \cite{MarMKBF02};
rapid rotation of the central black hole is not necessary in order
to explain the data in this case. Similar evidence for the broad
and skewed iron line has been discovered with different degree of
credibility in about eight other sources \cite{MarMKBF03}. In spite of
a limited similarity of the relativistic iron line in these sources,
we recall that recall that accretion discs have negligible gravity
in {\bh} candidates, and so we turn back to accretion discs galactic
centres.

Moving to a region upstream the accretion flow --- to the distance of
hundreds $\rs$ and farther out from the core of a galaxy,
the interaction between gaseous flows and other astronomical bodies 
(stars forming a nuclear
cluster) can substantially change the structure, accretion rate 
and chemical composition of the flow, so we briefly touch this
point, too. That way, the medium feeding the {\bh}
can be enriched above the level of usual solar metalicity 
\cite{ArtLW93,CheW99,ColH99,ColZ99a}. On
the other hand, moving satellites feel the presence of the interstellar
gas, especially in nuclei of active galaxies with high
concentrations of stars and the relatively dense interstellar environment
\cite{BegR78}. Accreting medium thus influences the motion and
evolution of these bodies. Even though self-gravitational effects of
the medium become progressively less important near the centre, 
a relatively accurate knowledge of their impact is desired in
order to derive, reliably, {\bh} parameters from rotational motion of
the surrounding luminous matter.
The dissipative interaction of stellar-mass satellites and the resulting
orbital decay proceed with a time-scale that is defined by properties of
both components of the system: in the inner disc, hydrodynamical drag
dominates the orbit evolution, while in the self-gravitating outer region,
the disc structure becomes clumpy and direct collisional interaction is
reduced. In even more remote parts, gravitational relaxation is
important for the cluster evolution, so the outer cluster is dynamically
distinct from the core.

Syer, Clarke \& Rees \cite{SyeCR91} proposed that a
stellar-mass body passing the galactic core can be captured on a 
quasi-stable bound orbit around a central super-massive {\bh}. 
In the simplest version of the model, stars have been
considered as bullets crashing on the disc surface
and passing through at hyper-sonic speed.
Zurek, Siemiginowska \& Colgate \cite{ZurSC94}, 
Armitage, Zurek \& Davies \cite{ArmZD96} and Ivanov,
Igumenshchev \& Novikov \cite{IvaIN98} examined 
various effects of passages (of individual stars)
through diluted gaseous environment (slab-disc geometry was assumed; 
other people explored different contexts 
of this problem \cite{CheW99,KarS01,NolK82,NovPP92}). 
Pineault \& Landry \cite{PinL94} discussed the statistical rate 
and distribution of impacts over the disc surface. 
Self-gravity of the disc medium was also considered 
\cite{Goo02,ShlB89,VokK98}, as well as
\gr\ effects in stellar motion \cite{KarV94,VokK93}, effects of the
orbital decay of satellites by gravitational radiation and close 
encounters \cite{HamKLA94,Nar00,QuiS87,SubK99}, 
or stellar coalescences \cite{Col67,FreB02,QuiS90,Rau99}.
Gradual changes of the physical and chemical state of satellites
and of the disc itself were discussed \cite{ArtLW93}. 
Radiation torques \cite{Ost83} as well as star--gas hydrodynamic interaction 
\cite{GolT80,Hag87,LinAW94} induce different channels of angular 
momentum transport in the medium. The clumpy structure \cite{FukHW00,Kum99}
of the outer, self-gravitating disc reduces
the efficiency of direct star--disc collisions, while turbulence in 
the medium tends to strengthen dissipation \cite{ZurSC94}.

Repetitive transitions bring the satellites into the plane of the disc at
almost circular orbits \cite{Rau95,SyeCR91}
(though, one can speculate of several effects that slow the process of
circularisation down, so that eccentric or non-equatorial orbits may be
maintained even at late stages \cite{GolS02,HolTT97,VilC02}).
Subsequent orbit evolution is reduced to a situation that was
addressed in connection with the formation and migration of bodies
inside proto-planetary discs \cite{NelPMK00,War97,YorBL93}. The inner
cluster becomes flattened and its structural dimension is
related to the size of the disc. One can immediately recognise plethora
of consequences connected with non-sphericity, anisotropy and satellite
segregation in the cluster: namely, the initial mass function is
modified, as well as the form of spectral-line profiles and stellar
velocity dispersions. The onset of star formation leads to gradual
conversion of gas into stars in the disc 
\cite{ColZ99,ColZ99a,HelS94}.

\begin{figure*}[tb]
\begin{center}
\includegraphics[height=0.37\textwidth]{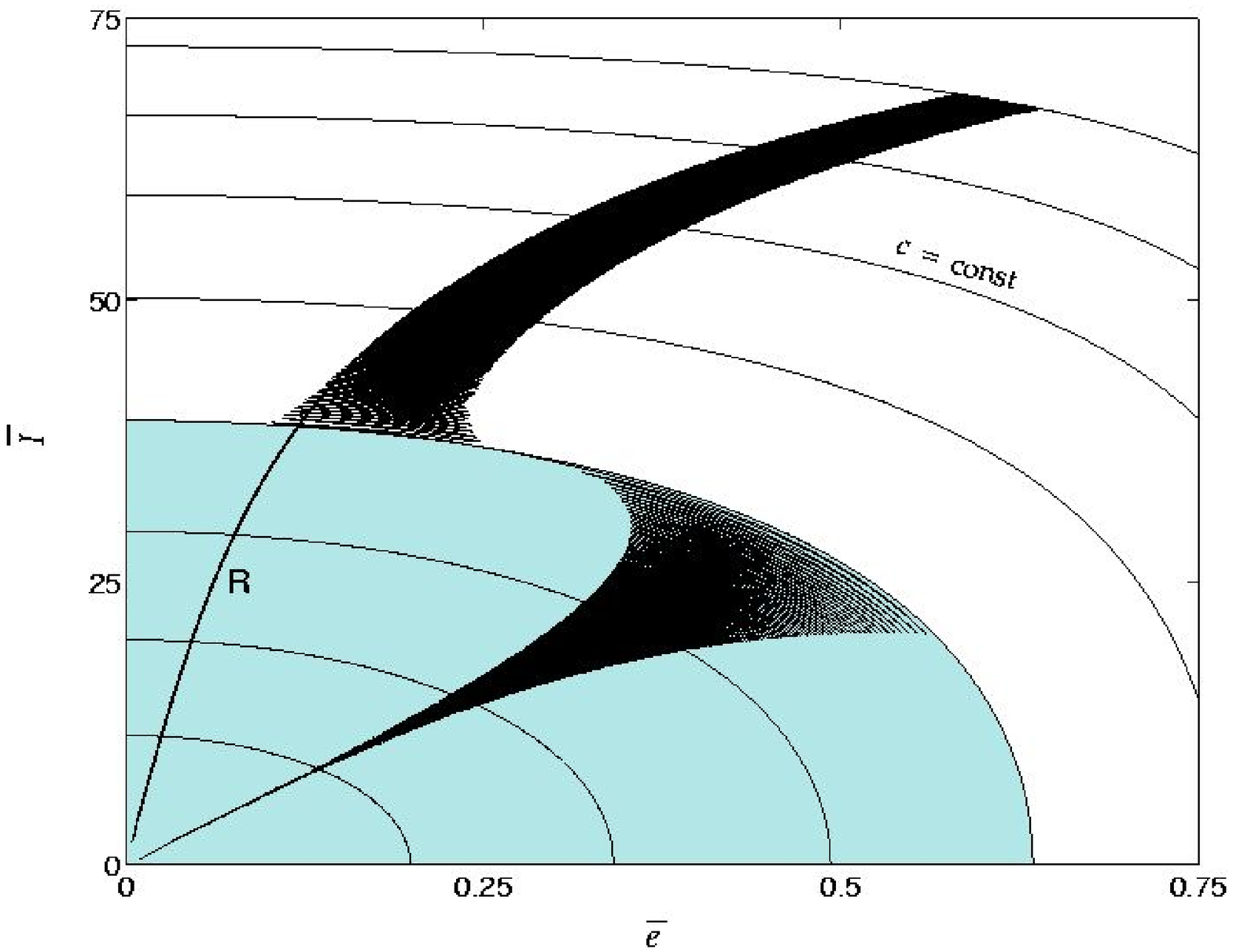}
\hspace*{2mm}
\includegraphics[height=0.37\textwidth]{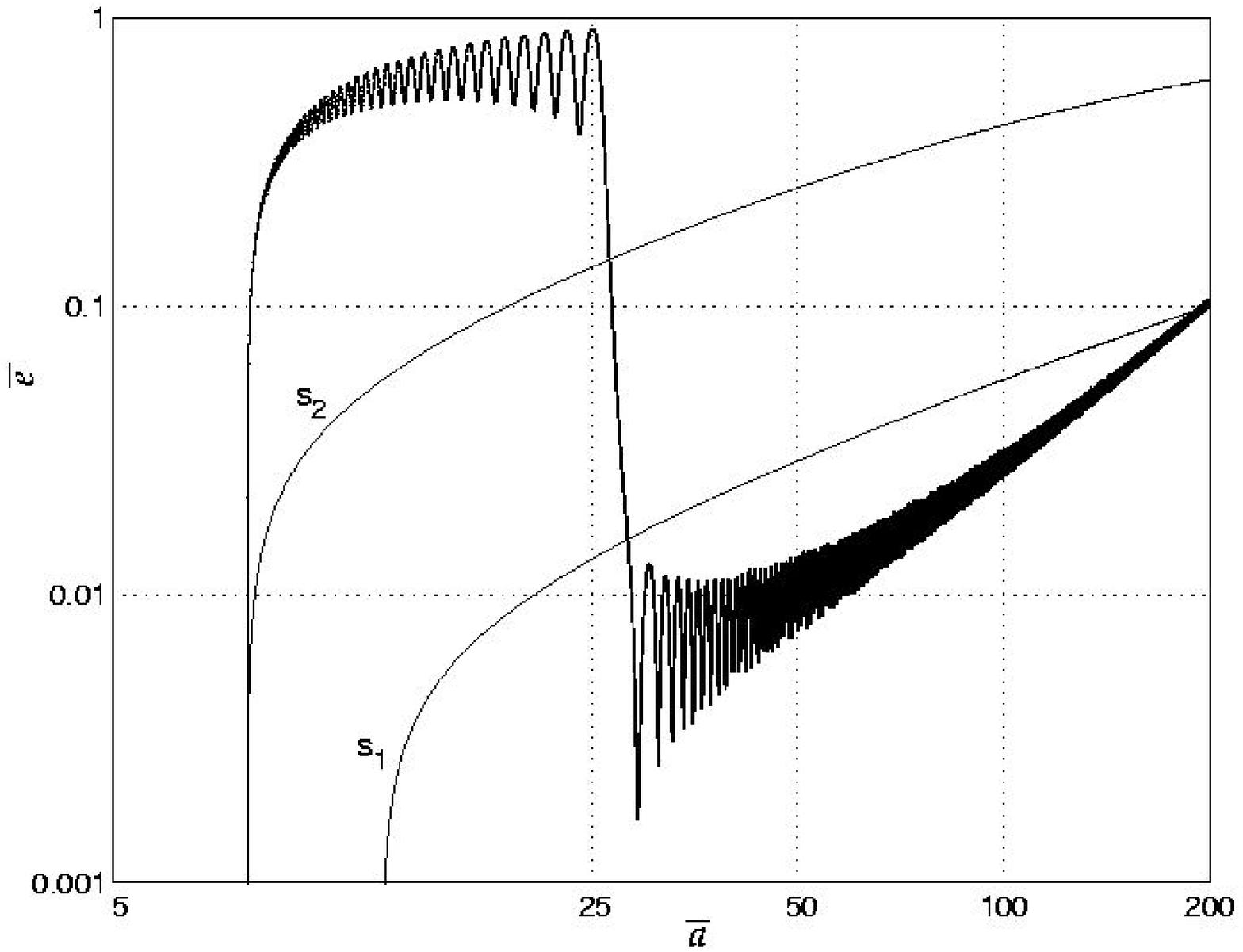}
\end{center}
\caption{Left: Mean inclination $\bar{I}$ vs.\ eccentricity $\bar{e}$
 of a satellite orbit that crosses a gravitating disc. 
 Contour lines of the Kozai parameter, $c=\mbox{const}$, are also plotted. 
 The orbital evolution consists of oscillations
 along $c$-contours and slow diffusion across them, towards the origin
 $\bar{e}=\bar{I}=0$. (This example has $\bar{e}_0=0.6$,
 $\bar{I}_0=70^{\rm{}o}$ as initial condition.)
 Kozai's quasi-integral is not isolating here, because of
 the orbital decay via dissipative transitions through the disc medium.
 An approximate analytic solution is also shown by a monotonic curve {\sf{}R},
 neglecting the influence of the disc gravity \cite{Rau95}.
 Right: Mean eccentricity $\bar{e}$ versus semi-major axis $\bar{a}$
 of a star crossing a gravitating disc. The orbit is represented 
 by an eccentricity curve with oscillations overlying the
 long-term decay. Also shown are two reference orbits having
 the same initial parameters (the solid curve ${\sf{}s_1}$)
 or the same final parameters (${\sf{}s_2}$) as a simplified 
 situation without the disc gravity (the latter two curves
 exhibit continuous and purely monotonic circularisation). 
 Large excursions in eccentricity occur if the disc mass 
 is non-zero, albeit it can be much less than the central mass
 (here, $\md\sim10^{-3}\mbh$). Figure adopted from~\cite{VokK98}.}
 \label{fig:vk}
\end{figure*}

Star-burst phenomenon has been indeed identified in some active galactic 
nuclei on $\sim10^2\;$pc scales, which could represent outskirts of a dusty 
torus surrounding the central black hole \cite{Are03,SchiET01}. 
Indeed, violent star formation could coincide with the outer parts of a 
massive nuclear cluster that is embedded within the dusty torus.
Even if $\md\ll\mbh$, the effect of the disc gravity on 
the long-term evolution of stellar orbits should be taken into account
together with dissipation via star--disc (hydrodynamical) interactions. 
This stands in contrast with the fact that gravitation of accretion 
discs has been neglected in most works exploring the nuclear cluster 
structure. The two influences are linked through their dependence 
on the disc density. A remark 
is worth making regarding the mass ratio $q$ (which
was criticised as unsuitable self-gravity
criterion in section~\ref{sec:weak}): it is a useful measure
to assess the importance of stellar transitions across a self-gravitating
disc in galactic nuclei, because attraction of the disc and the effects of
transitions are both increasing, roughly proportionally, with
$\md$. 

Circularisation of the orbits, evolution of their inclination and
the resulting capture rate are visibly affected by the disc gravity. 
An important point is the range of time-scales: (i)~dynamical
period (of satellite stars revolution around the central mass;
the shortest time-scale), (ii)~period of oscillations in orbital elements
(medium time-scale; these oscillations are due to disc gravity) and
(iii)~time for bringing the orbital plane into the disc (caused by 
successive interactions with the disc; long time-scale).
The orbital decay continues also when satellites are embedded
in the disc plane; its rate varies depending on the ability
of the body to clear a gap along its trajectory through the medium.
Different modes depend on the satellite masses, sizes and on the disc 
thickness and surface density at corresponding 
radius. The evolution of orbital parameters is an issue
in extra-solar planetary research \cite{Per00}, which
tackles similar systems on much small scales --- cf.\ 
the discussion of planetary formation and migration
\cite{ArmLLP02,HolTT97,LinI97,NelPMK00}. These topics
have not yet been pursued to much detail in the context of galactic 
nuclear clusters embedded in the gaseous environment.

One can understand the origin of orbital parameters oscillations in 
terms of Kozai's theory \cite{Koz62}, taking into account the gradual 
adiabatic diffusion of quasi-integrals of motion. 
In order to see the reason for orbital oscillations,
one borrows from classical mechanics' traditional approach to
the disc gravitational potential $\Psid(\vec{r})$ as a
perturbation acting on the satellite motion
\cite{BroC61,LemD91,Mor02}. Again, the
total potential is supposed to be dominated by the central field of 
the core, while the disc contribution 
${\Psid}$ can be transformed by a canonical transformation to the 
form which depends on mean orbital elements 
(eccentricity $\bar{e}$, inclination $\bar{I}$, and argument 
of pericentre $\bar{\omega}$; the bar denotes mean values) and does
{\it{}not} depend on the fast-changing variable (mean anomaly). 
See \cite{VokK98} for the explicit form of this new $\barPsid$ in terms of
integration (over the mean anomaly) along an unperturbed Keplerian
ellipse; in that paper, Kuzmin's potential was employed as a simple
model for the disc gravitational field (it would be quite
interesting to examine a generalization in terms of Miyamoto \&
Nagai's potential \cite{MiyN75}, which can be still performed 
analytically). In the first order of perturbation
theory one finds that ${\barPsid}\sim{\rm{const}}\equiv{\barPsidn}$; 
the mean semi-major axis $\bar{a}$ is also constant and
differs from the actual osculating semi-major axis by short-period terms
only. The mentioned averaging procedure allows us to disregard short-term
effects and concentrate on the long-term evolution. 
Furthermore, owing to the axial symmetry of $\Psid(R,z)$
exists an additional constant of motion \cite{Koz62}:
$\sqrt{1-\bar{e}^2}\cos\bar{I}=c\equiv\mbox{const}$
(figure~\ref{fig:vk} left). This way the problem of orbital evolution is 
reduced to the evolution of ${\bar{e}}$ and
${\bar{\omega}}$ which are further constrained by condition
${\barPsid}\left({\bar{e}},{\bar{\omega}};c,{\bar{a}}\right)=
{\barPsidn}$. Contours of the perturbation potential ${\barPsid}$ 
provide a convenient representation of mean
orbital elements in the $(\bar{e},\bar{\omega})$-plane.
Finally, {\it{}with the inclusion of orbital dissipation
in the gaseous medium, one finds that an adiabatic drift occurs and drives
the orbit across the contours of constant $\barPsid$;
subsequently, the drift provokes excursions of the orbital 
elements when a separatrix is traversed}. A representative 
example of a trajectory is shown in figure~\ref{fig:vk} 
(right). The interplay between
Kozai's mechanism and gradual changes of the perturbation potential
can give occasional (but large) eccentricity jumps; one of them is clearly 
visible in the figure.

\begin{figure*}[tb]
\begin{center}
\includegraphics[width=0.93\textwidth]{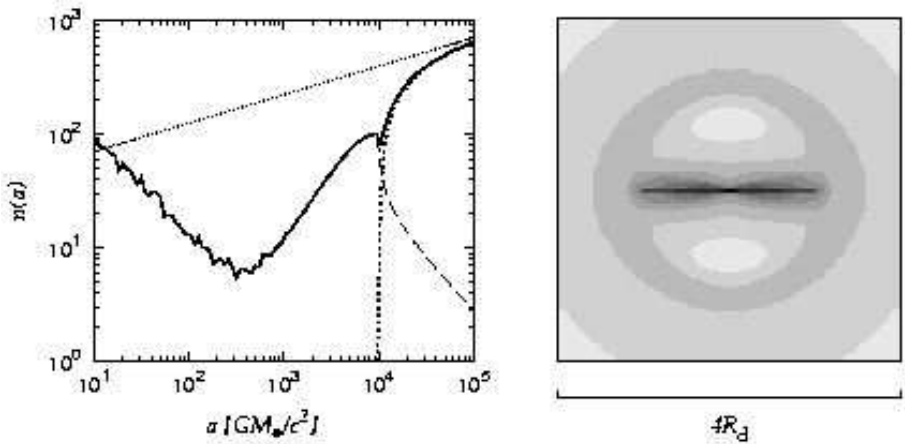}
\end{center}
\caption{Structure of a nuclear cluster in interaction with a planar, 
 gravitating disc. In this model \cite{KarSS02,Sub01}, the
 long-term evolution of stellar orbits is influenced by dissipative
 transitions across the disc slab and by (small) gravitational attraction 
 of the disc itself.
 Left: Number density $n_{\ast}(a)$ of satellites within a shell of
 semi-major axis ${\langle}a,a+{\rd}a\rangle$ (solid curve). Two populations
 can be distinguished, corresponding to the satellites of the outer cluster
 (dotted) and to the highly flattened inner cluster (dashed). Thin dotted
 power-law line indicates the equilibrium structure of the Bahcall-Wolf
 cluster ($n\,\propto\,a^{1/4}$).
 Right: The number density $n_{\ast}(a)$ is represented by different levels of 
 shading here. An azimuthal section of the cluster is plotted. 
 In both plots, the gradual concentration of satellites towards the
 equatorial plane and formation of a ring structure are evident.}
\label{fig:cluster}
\end{figure*}

Next step would be to consider a consistent scheme for the whole cluster of 
satellites. In their classical paper, Bahcall \& Wolf \cite{BahW76}
anticipated a way to reveal a hypothetical central massive hole
(embedded in a collisional environment of a globular cluster) 
through stellar velocity dispersion and
broadened wings of line profiles integrated over the {\bh} domain of
influence. In order to calculate the form of central cusps, the authors
employed time-dependent diffusion equation. Of other assumptions, the
most important was the spherical symmetry and isotropy of the
cluster itself. Probability density of satellites can be expressed in terms
of their semi-major axes: $n_{\ast}(a)\,\propto\,a^{1/4}$, where
$n_{\ast}(a)\,{\rd}a$ denotes the number of satellites in a shell with
radius $a$ and width ${\rd}a$. Corresponding spatial density
$n_{\ast}(r)$ is thus proportional to $r^{-7/4}$. Modifying 
the model assumptions (namely, that about the importance of 
collisions) leads to different but similar 
density profiles (power-law at large radii). For example, 
the well-known Plummer \cite{Plu11} sphere has a 
finite-density core with density falling off as $r^{-5}$, 
while Jaffe \cite{Jaf83} and Hernquist \cite{Her90} models both 
decline $\propto\,r^{-4}$ at large radii. The latter case has a 
sound theoretical basis in the mechanics of violent relaxation. 
See also \cite{Kin66,Pee72,ShaM78,You80} and, for
more recent discussion and references, see \cite{Ale99,TakL00}. 
Further, the collisional steady-state solutions have been
reconsidered as a model for a cusp in the center of a galaxy
\cite{EvaC97}.
Conclusions of these papers can be compared with more advanced
models that relax some of the assumptions and include new effects,
e.g.\ resonances, enhanced loss-cone depletion or rotation of the cluster
\cite{FraR76,KimE02,QuiHS95,RauT96,ZhaHR02}. 
Nevertheless, self-consistent treatment of non-spherical stellar systems
with the effects of self-gravity {\it{}and} the presence of an accretion 
disc are still an open problem. The disc
induced dissipative orbital decay is distinct in its tendency to 
flatten the stellar component and even form a ring of stars that
could be resolved observationally. 

Figure~\ref{fig:cluster} illustrates a stationary density
profile of the cluster in a model where its members revolve around
a {\bh} and their orbits slowly decay due to dissipative 
effects.
One can distinguish different populations of satellites
(see the left panel): those which form almost spherical cluster
far from the centre, outside the disc outer radius, and those
which have been dragged into the disc plane near the centre 
(the latter form a ring). Hence it is a certain generalisation 
of the Bahcall--Wolf \cite{BahW76} cluster. A
complementary representation is provided in the right panel of 
figure \ref{fig:cluster} which shows an azimuthal section across the 
cluster. Again, the inner cluster is evidently flattened and its dimension 
is tightly related to the size of the disc. It should be recalled that,
with this approach, a relatively small volume 
is explored near {\bh}, where the central mass has
direct dynamical influence on the embedded stars and gas
(in other words, rotational support of the stellar motion is not
a negligible fraction compared to their random motions). 
The linear size of this region is a function of $\mbh$ and $\rmsv$, 
and is typically less than $\sim10\,$pc.

Note that the $\log\mbh\sim4\log\rmsv$ relation resembles the 
empirical correlation which has been determined in a large sample 
of galactic bulges \cite{FerM00,Gebetal00} (although the constant
of proportionality can be different in the innermost region around 
the central {\bh}, which is unresolved by current observations). 
Hence, in the inner cluster of stars, the obtained velocity dispersion 
should be determined by the joint action of gravity and of other agents, 
including dissipation in the interstellar environment.
It was proposed \cite{KarSS02,Sub01} that the effect of interaction 
between stars and the accretion 
flow shapes the observed $\mbh$--$\rmsv$ relation in 
the innermost region of some active galaxies 
(those with sufficiently dense accretion discs in the 
core). The model of star--disc interaction leads to
an axially symmetric, but non-spherical central cluster, and so
the exact value of the power-law index of the $\mbh$--$\rmsv$ relation
depends also on the observer line of sight (the expected central 
velocity dispersion comes out somewhat larger for
systems seen edge-on than for pole-on view).
The model therefore represents rather extreme case, however, it
is obvious that the interaction between stars and
the intergalactic environment should be taken into account 
also at larger distances from the core \cite{WozCEF03}.
Miralda-Escud\'e \& Kollmeier \cite{MirK03} applied a
similar idea to a much wider region (and hence, applicable to 
actual present-day observations). These authors propose star-disc 
collisions are a self-regulating process that helps to feed the central {\bh}
and control its growth. The influence of the collisions remains imprinted 
in the bulge mass--stellar velocity dispersion relation, which concerns
the region in galactic nuclei greatly exceeding the domain
of direct gravitational influence of the super-massive {\bh}.

Naively, one could expect that dissipation of satellite's orbital 
energy and momentum should always lead to a very dense nuclear 
cluster. However, the outcome
may be more complicated because of non-sphericity and satellite
segregation (caused by differing pace of orbital evolution, according 
to the mass function and the distribution of sizes of stellar bodies). 
For example, recently, Zier \& Biermann
have showed \cite{ZieB01}, on the basis of numerical simulations, that a
flattened shape of the cluster may arise due purely to
gravitational effect of a massive binary {\bh} in the core. They found
that a stellar ring is formed on a parsec scale in the orbital
plane of the central binary, while the cluster is gradually depleted.
Another approach to (self-gravitating, eccentric) galactic stellar discs
was applied to the double nucleus of M\,31 \cite{SalS01} and
even the case of Milky Way was discussed recently \cite{LevB03}.
Indeed, as indicated also in this paper, 
there is a broad variety of reasons why the compact nuclear cluster 
should acquire a non-spherical structure in the course of evolution
\cite{Lauetal02,SchiET01}. The effects exhibited by self-gravitating 
accretion flows are merely one ingredient of a complex problem. 

\section{Conclusions}
Gravitational field of black-hole discs induces great diversity  of
effects and gives ample scope for new ideas, of which we could  touch
only very limited section. First of all, as noted in the first part of
this paper, even some very basic questions remain still open within the
framework of exact general relativistic solutions for the black
hole--disc gravitational field. Our interest in global gravitational
effects connected with astrophysical discs has lead us to focus on
systems around super-massive {\bh}s that reside in nuclei of galaxies,
in which case the onset of self-gravity appears inevitable. Recent ideas
of heavy tori being formed in certain moments of neutron-star binary
coalescence make these topics fitting also to stellar mass objects,
although their ``discs'' must be very different in many respects.
Because of expected instabilities, accreting {\bh}s are pertinent also
for forthcoming gravitational wave experiments. 

Astrophysical black holes are supposed to be embedded in complex cosmic
environment which consists of multiple sub-systems spanning enormous
range in the parameter space of masses, lengths and time-scales. 
Gaseous discs represent just one of the components in this formation.
Under such circumstances, analytical means can offer insight into
physical mechanisms which are connected with the disc, but specific 
analytical solutions lack generality and must be inevitably simplified.
In recent decades, much effort has been devoted to clarify, which
processes dominate the evolution and what are the applicable ranges of
model parameters. It is quite possible that some regimes of the disc
operation have been missed in previous works.

In this respect, an attractive recent speculation suggests that
intermediate mass black holes (with $\mbh\sim10^3M_{\sun}$) might also
exist, possibly in cores of some globular clusters. If this new family
of black holes is revealed and if they also accrete gas from their close
environs (current observational evidence is rather scarce), it would
provide a fascinating opportunity to study accreting black hole sources
for which the dynamical time is of the order of $\sim1$~second. In
other words, the expected variability of such object could span over
intervals that can be readily sampled over many periods with available
technology.

We concentrated our attention on a fairly conventional 
case of gas discs around super-massive black holes at small, 
sub-parsec length-scales where the motion of stellar-mass orbiters can
serve as an accurate tool to probe the system. This idea requires
various intervening influences (such as the perturbation of the central
gravitational field caused by the disc/torus own gravity, or the
orbital dissipation in the gaseous interstellar environment) to be
understood with sufficient accuracy. Interesting and fundamental
questions, namely, whether and where exactly the onset of self-gravity
occurs in real accretion discs, still remain open to further research.

\ack
We are grateful to J~Miller for very helpful discussions and
criticism of the first version of the manuscript. We thank J~Bi\v{c}\'ak
for his support, in various ways, of this work. We also very much
appreciate comments and advice on different points from M~Abramowicz,
S~Collin, B~Czerny, P~Harmanec, A~Martocchia, G~Matt, J~Palou\v{s},
L~Rezzolla, L~\v{S}ubr, D~Vokrouhlick\'y and participants of the 
Prague relativity seminar. We 
acknowledge support under grants GACR 202/02/0735, GACR 205/03/0902 and
GAUK 188/2001. Part of this paper was written during VK's visit to the
UKAFF at the University of Leicester.

\end{document}